\documentclass[useAMS,usenatbib]{mn2e}

\usepackage{graphicx}
\usepackage{url}
\usepackage{natbib}
\usepackage{my_aas_macros}
\usepackage{caption}
\usepackage{array}
\usepackage{multicol}
\usepackage{amsmath}

\newcolumntype{C}[1]{>{\centering}m{#1}}

\title[Stellar activity in Blanco-1]{A search for flares and mass ejections on young late-type stars in the open cluster Blanco-1\thanks{Based on observations made with ESO Telescopes at the Paranal Observatory under programme ID 089.D-0713(B).}}
\author[M. Leitzinger et al.]{M. Leitzinger$^{1}$\thanks{E-mail: martin.leitzinger@uni-graz.at}, P. Odert$^{1}$, R. Greimel$^{1}$, H. Korhonen$^{2, 3}$, E.W. Guenther$^{4}$, \newauthor A. Hanslmeier$^{1}$, H. Lammer$^{5}$, M.L. Khodachenko$^{5}$ \\
$^{1}$Institute of Physics, Department for Geophysics, Astrophysics, and Meteorology, NAWI Graz, Universit\"atsplatz 5, 8010 Graz, Austria\\
$^{2}$Finnish Centre for Astronomy with ESO (FINCA), University of Turku, V{\"a}is{\"a}l{\"a}ntie 20, FI-21500 Piikki{\"o}, Finland\\
$^{3}$Niels Bohr Institute, University of Copenhagen, Juliane Maries Vej 30, DK-2100 Copenhagen, Denmark\\
$^{4}$Th\"uringer Landessternwarte Tautenburg, 07778 Tautenburg, Germany\\
$^{5}$Space Research Institute, Austrian Academy of Sciences, Schmiedlstra\ss{}e 6, 8042, Graz, Austria\\
}

\begin{document}

\date{Accepted 1988 December 15. Received 1988 December 14; in original form 1988 October 11}

\pagerange{\pageref{firstpage}--\pageref{lastpage}} \pubyear{2002}

\maketitle

\label{firstpage}

\begin{abstract}
We present a search for stellar activity (flares and mass ejections) in a sample of 28 stars in the young open cluster Blanco-1. We use optical spectra obtained with ESO's VIMOS multi-object spectrograph installed on the VLT. From the total observing time of $\sim$ 5 hours, we find four H$\alpha$ flares but no distinct indication of coronal mass ejections (CMEs) on the investigated dK-dM stars. Two flares show ``dips'' in their light-curves right before their impulsive phases which are similar to previous discoveries in photometric light-curves of active dMe stars. We estimate an upper limit of $<$4 CMEs per day per star and discuss this result with respect to a semi-empirical estimation of the CME rate of main-sequence stars. We find that we should have detected at least one CME per star with a mass of 1-15$\times10^{16}$g depending on the star's X-ray luminosity, but the estimated H$\alpha$ fluxes associated with these masses are below the detection limit of our observations. We conclude that the parameter which mainly influences the detection of stellar CMEs using the method of Doppler-shifted emission caused by moving plasma is not the spectral resolution or velocity but the flux or mass of the CME.
\end{abstract}

\begin{keywords}
stars: activity -- stars: flare -- stars: late-type -- open clusters and associations: individual: Blanco-1.
\end{keywords}

\section{Introduction}

Coronal Mass Ejections (CMEs) represent the most energetic activity phenomenon of the Sun. In the most energetic events masses in the range of 10$^{16}$-10$^{17}$g  \citep{Vourlidas2010} are expelled into the heliosphere with velocities up to 3000~km~s$^{-1}$. 
These highly energetic particles ejected from the Sun and solar-like stars can interact with the atmospheres of orbiting planets. The planetary magnetic field may shield the planetary atmosphere against charged particles, but if the planetary magnetic field is weak and/or if the host stars' X-ray and EUV (XUV) radiation heats the upper planetary atmospheric layers so that they expand beyond the planetary magnetic field, these layers are not protected by the planetary magnetic field anymore. The incoming stellar XUV radiation ionizes the atmospheric neutral gas which interacts with the charged particles of CMEs and the solar/stellar wind. The planetary atmosphere may get eroded \citep{Lammer2007, Khodachenko2007, Cohen2011, Lecavelier2012, Khodachenko2014} if the planet orbits its host star at close distances, as it is the case for planets orbiting dM stars in the habitable zone which is located at $\sim$0.2~AU \citep{Kopparapu2013}. Therefore, observationally deduced CME rates of late-type main- or even pre-main sequence stars are needed to yield the required input for the above mentioned studies which investigate the influence of the stellar plasma environment on exo-planetary atmospheres.\\
Solar CMEs are well investigated in contrast to stellar CMEs because the methods of detection differ. Solar CMEs are monitored mainly in white light coronagraph images, such as the Large Angle and Spectrometric Coronagraph Experiment (LASCO) onboard the Solar and Heliospheric Observatory (SOHO) and the Solar TErrestrial RElations Observatory (STEREO). On stars such a technique is not feasible because they appear as point sources in our telescopes and the expected emission from stellar CMEs is too small in comparison with the stellar luminosity. The only way to obtain information on stellar CMEs is by using other observational signatures known from solar CMEs. Statistical investigations of solar CME parameters yield mean CME masses of $\sim$10$^{14}$g, mean velocities of 500-600~km~s$^{-1}$, and mean kinetic energies of  $\sim$10$^{29}$erg. Those values represent mean values over a full solar cycle \citep{Schwenn2006, Vourlidas2010, Webb2012}.\\ 
On the Sun flares and CMEs are related phenomena probably caused by the same physical process. Solar energetic flares (X-class) and CMEs show a high correlation, whereas less energetic CMEs and flares show a lower one \citep{Yashiro2009}. Therefore the occurrence of stellar energetic flares may be used as an indicator for stellar CME occurrence, at least for solar-like stars, relying on the solar-stellar analogy \citep{Aarnio2012, Drake2013}. However, there is still a lack of observational confirmation that such relations can be extrapolated to young and frequently flaring stars.\\
Stellar flares follow a power law (see section~\ref{sectionCMErate}, Eq.~\ref{eq1}) similar to solar flares where the slope varies within -1.5 ... -2.5 \citep{Audard1999, Crosby1993, Audard2000, Veronig2002, Aschwanden2012}. 
\citet{Audard2000} derived a power law for flare rates from young and active stars depending on their X-ray luminosities. Using the solar X-ray luminosity \citep[10$^{26.8}$ - 10$^{27.9}$ erg~s$^{-1}$, ][]{Judge2003} we calculate 0.1-0.6 flares with energies $>$ 10$^{32}$ erg per day. Using the X-ray luminosity of the young solar analog EK Dra  \citep[log$L_{X}$=29.9~erg~s$^{-1}$, ][]{Guedel1995} we calculate 48 flares with energies $>$ 10$^{32}$ erg per day. From these calculations one can see that X-ray luminous young stars may exhibit an enormous number of energetic flares in comparison to our Sun.\\
For stellar flares detected in H$\alpha$ the energies are high. On the Sun, H$\alpha$ flares are not seen in full-disk integrated light. For the active dMe star AD~Leo, \citet{CrespoChacon2006} find a flare-rate from chromospheric lines of $>0.71$ flares/h. Two flares were also observed in H$\alpha$ with energies of $1.34\times10^{30}$ and $7.5\times10^{29}$~erg, respectively. \citet{Hawley2003} studied several flares on the same star and derived H$\alpha$ energies of $4\times10^{29}-3.75\times10^{30}$~erg and a range of total (optical-UV) energies of $3.75\times10^{31}-2.7\times10^{32}$~erg. \citet{Kowalski2013} found H$\alpha$ flare energies of $5\times10^{29}-1\times10^{32}$~erg for several nearby active M dwarfs.\\
Signatures of solar CMEs that may be applied to stellar observations can be found in different wavelength ranges. In the meter, decameter, and even in the hectometer wavelength range solar CMEs show a strong correlation to radio type II bursts \citep{Reiner2001a, Gopalswamy2001}. Type II bursts have been recognised as signatures of shock waves. CMEs are possible drivers for shock waves in the solar atmosphere. 
In the past decades there have been several attempts to detect stellar radio bursts in the decameter regime \citep{Jackson1990, AbdulAziz1995, Abranin1998, Leitzinger2009}, but a stellar analog of the solar type II burst has not been detected yet.\\
On the other side of the electromagnetic spectrum CMEs are correlated with regions of decreased emission known as ``dimmings'' seen at X-ray and EUV wavelengths representing the early stage of a CME.  Dimmings are known to last for hours and occur in regions of either plasma evacuation or changing temperature \citep{Gopalswamy2000}. On the Sun more than 80\% of CMEs can be tracked back to dimming regions \citep{Bewsher2008}. On stars, X-ray dimmings have been detected on the binary V471~Tau suggested to be coronal loops anchored to the dK component \citep{Jensen1986}. \citet{Haisch1983} detect an energetic X-ray flare on Proxima Cen which they suggest to be ``two-ribbon'' like. Furthermore they find an X-ray deficit which these authors interpret as prominence eruption. In UV light curves, \citet{Schroeder1983} find transient absorption features caused by a large cloud or prominence. Suddenly appearing absorption lines were detected on V471~Tau \citep{Guinan1986} which are likely to be cool coronal loops.\\
A more direct signature of moving plasma are Doppler-shifted emission or absorption features of spectral lines. Signatures of solar eruptive prominences/CMEs have been so far detected by \citet{Den1993}, who investigated an M-class flare using H$\alpha$ filter- and spectrograms. A strong mass ejection was detected in the spectrograms, with radial velocities of ∼600~km~s$^{-1}$ . The spectrograms showed a distinct blue-wing absorption as a signature of ejected, relatively cool solar plasma. \citet{Ding2003} performed a multi-wavelength study of a two-ribbon flare and its associated filament eruption. These authors detected a blue-wing absorption feature in H$\alpha$ caused by the erupting filament, with a line of sight velocity of 210~km~s$^{-1}$. \\
Also on stars spectral signatures related to CMEs have been detected. \citet{Houdebine1990} presented spectroscopic observations of the young ($\sim$200~Myr) and active dMe star AD Leo. A spectacular enhancement occurred in the blue wing of the H$\gamma$ line. During four consecutive spectra($\Delta$ t $\sim$240s) the evolution of this enhancement could be followed. The authors deduced a projected velocity of 5800~km~s$^{-1}$ and estimated an ejected mass of $\sim$8$\times$10$^{17}$g. The fastest solar CMEs reach velocities of up to $\sim$ 3000~km~s$^{-1}$ and the most massive solar CMEs reach values in the order of 10$^{17}$g. So the velocity is twice as high and the mass is almost an order of magnitude higher. \citet{Guenther1997} monitored a sample of classical (CTTS) and weak-line T-Tauri stars (WTTS) in the Taurus Auriga region and detected an extra-emission in the blue wing of H$\alpha$ on a WTTS which was clearly identified as mass ejected from the star. The authors deduced a projected velocity of 600~km~s$^{-1}$ and estimated a mass in the range of 10$^{18}$ - 10$^{19}$g, which is a factor 10-100 higher than the most massive solar CMEs. 
\citet{FuhrmeisterSchmitt2004} analysed optical spectra of the 9~Gyr old dMe star DENIS 104814.7-395606.1 and found an enhanced blue wing seen in H$\alpha$ and H$\beta$. The authors deduced a projected velocity of 100~km~s$^{-1}$ and preferred in their conclusion a ``dynamic scenario'' which corresponds to a mass ejection event rather than a ``static scenario'' corresponding to a plasma cloud co-rotating with the star.\\
Extra-emissions/absorptions in UV/FUV were also interpreted as stellar CMEs. \citet{Bond2001} analysed Ly$\alpha$ spectra of the eclipsing binary V471~Tau consisting of a dK2 star and a white dwarf. The Si\,{\sc iii} (1206\AA) line located close to Ly$\alpha$ was affected by transient absorptions lasting for several exposures. The authors interpreted those as `` ... CMEs from the K2 dwarf pass across the line of sight to the white dwarf.'' From the number of detected transients and the total observing time, as well as considering the geometry of the system, the authors deduced a CME rate of 100-500 per day. In comparison, the Sun shows about 4 CMEs per day in solar maximum \citep{Yashiro2004}. \citet{Leitzinger2011a} analysed a time series of FUV spectra of the dMe star AD Leo and detected an extra emission in the blue wing of the first component of O\,{\sc vi}  duplet at 1032\AA. The authors could exclude a co-rotating plasma cloud \citep[cf.][]{FuhrmeisterSchmitt2004}, but the low projected velocity of 84~km~s$^{-1}$ could not be unambiguously interpreted as a mass ejection event, although prior to the extra-emission a flare occurred. The temporal separation of the flare and the extra-emission was about 2 hours.\\
In the present study we aim to investigate stellar activity in form of flares and CMEs of young stars in the open cluster Blanco-1 using H$\alpha$ spectroscopy. In section~2 we describe the observations, in section~3 we present the results, in section~4 we discuss our findings, and present our conclusions in section~5.
\begin{figure}
\begin{center}
\vspace*{0cm}
\includegraphics[width=\columnwidth]{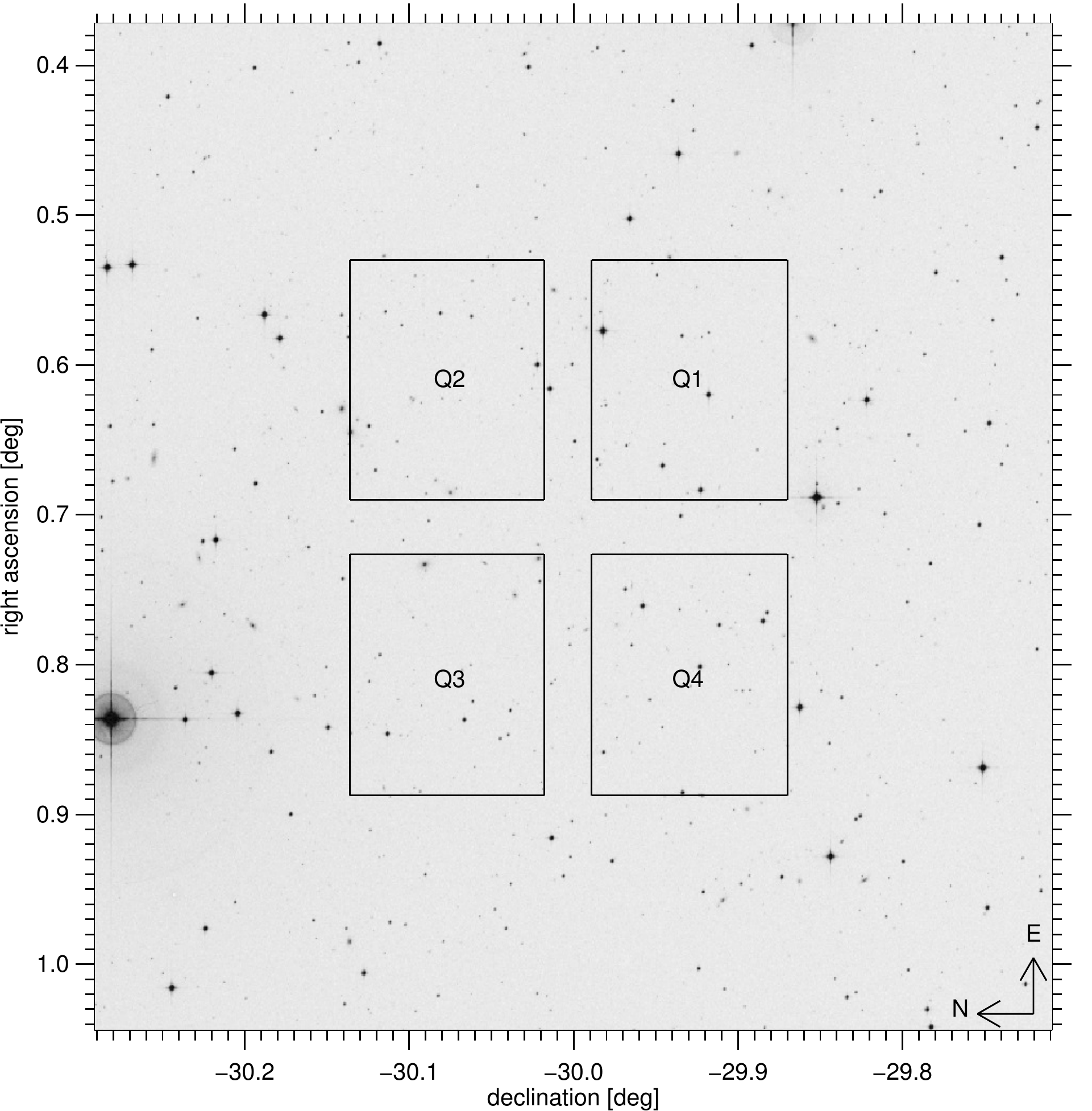} 
 \caption{The four quadrants of VIMOS overlayed on a digitized sky survey (DSS) image of Blanco-1.}
\label{FoV}
 \end{center}
 \end{figure} 

\section{Observations}
\label{observations}
To optimize the detection of stellar flares and CMEs we used multi-object spectroscopic observations of coeval stars of young open clusters. The VIsible Multi Object Spectrograph \citep[VIMOS, ][]{LeFevre2003} mounted on Nasmyth focus of the Very Large Telescope (VLT) of the European Southern Observatory (ESO) located at Cerro Paranal in Chile turned out to be an ideal instrument for the investigation of stellar CMEs. To get the maximum number of stars into the field of view (FoV) which have a similar age we select coeval cluster stars. As we are interested in young and active stars we concentrate on young open clusters.\\ 
We selected the multi object spectroscopy (MOS) mode which has four quadrants with a FoV of 7'$\times$8' each, and the high resolution grism (HR) orange with default filter GG435. The wavelength coverage for this setting is 520-760~nm, the dispersion is 0.6\AA{} per pixel, and the spectral resolving power is 2150. The requested observing time was 10.85 hours including pre-imaging, read-out time, and calibration. The net on-source observing time was 4.95 hours. The 11 observing blocks (OBs) were executed in service mode between 3 July and 8 August 2012. Each OB is about one hour in length and contains nine spectra of which each spectrum has an integration time of three minutes.\\
The nearby \citep[$\sim$~200 -240~pc, ][]{Epstein1985, vanLeeuven2009} and young (30-145~Myr, see section~\ref{Halphaprofiles}) open cluster Blanco-1 \citep[$E(B-V)$=0.016, ][]{Kharchenko2005}, which spans an area of several square degrees and lies at a high galactic latitude, turned out to be an ideal target, because of its age, proximity, and well studied parameters of the cluster stars.
\renewcommand\thefootnote{\alph{footnote}} 
\begin{table*}
 \centering
 \begin{minipage}{180mm}
  \caption{Observed stars with a S/N $>$ 20 in Blanco-1. Apparent $J$ and $K$ magnitudes were taken from 2MASS, NUV GALEX magnitudes (AB system) were taken from GALEX GR7, and apparent $V$ magnitudes were taken from \citet{Mermilliod2008, Platais2011}. $(B-V)_{0}$ indices were taken from \citet{PanagiOdell1997, Pillitteri2003, Jackson2012} or computed using an $E(B-V)$ value of 0.01 taken from \citet{Kharchenko2005}. X-ray luminosities were taken from \citet{PanagiOdell1997, Pillitteri2003, Jackson2012} or computed from fluxes measured by the X-ray Multi-Mirror-Newton (XMM-Newton) satellite \citep{Watson2009}. Masses and temperatures were taken from \citet{Pillitteri2003}, and membership probabilities from \citet{Platais2011}. Spectral types and luminosity classes were taken from the present study. The last two columns indicate if H$\alpha$ is in emission (em) or absorption (abs) and the observed number of flares (in brackets the abbreviations of the flaring stars of this study, as given in section~\ref{subflare}, are noted). The quadrants correspond to the quadrants in Fig.~\ref{FoV}}.
  \label{bigtable}
  \begin{tabular}{ccC{1.1cm}ccccccccc}
  \hline
   
      2MASS ID    &   $J$  & $K_{S}$ &  NUV  &  $V$   & $(B-V)_{0}$ & log$L_{X}$     & mem  &  spectral  & luminosity & H$\alpha$ & flare  \\
                  &  [mag] &  [mag]  & [mag] & [mag]  &    [mag]    & [erg~s$^{-1}$] & [\%] &    type    &   class    &           &            \\
    \hline
    quadrant 1\\
    \hline
 00032088-3004472 & 13.901 & 13.075  &  -    & 18.16  &    -        & 28.71          &    3     & M4-5  &      V     &  em        &  1 (FS1)    \\
 00032107-3004040\footnote{In a colour magnitude diagram ($V-J$, $J$) these stars are significantly below the cluster sequence and therefore confirm the non-membership as given in \citet{Platais2011}. } & 13.093 & 12.456  &  -    & 14.91  &    -   &   -                  &    0     & K5    &      V     &  abs       &   0    \\
 00030027-3003215 & 11.162 & 10.555  & 19.59 & 12.89  &  0.91  & 29.62                &    93    & K4-7 &      V     &  abs       &   0    \\
 00033082-3003091\footnotemark[1] & 14.379 & 14.088  & 20.18 & 15.56  &  -     &   -                  &    0     & K3-4  &    III/V   &  abs       &   0    \\
 00030351-3002024\footnotemark[1] & 14.453 & 14.211  & 18.45 & 15.46  &  -     &   -                  &    0     & K3-4  &    III/V   &  abs       &   0    \\
      \hline
   quadrant 2\\
     \hline
 00031154-2958103 & 14.299 & 13.473  &   -   & 18.51  & 1.50   & 29.16                &    23   & M4-5  &     V      &  em        &  1 (FS2)   \\
 00032417-2956229 & 12.493 & 11.727  &   -   & 14.95  & 1.30   & 28.77                &    91    & K7-M0 &     V      &no H$\alpha$&   0    \\
 00032466-2955146 & 12.172 & 11.480  & 21.74 & 14.18  & 1.12   & 29.32                &    92    & K5-7 &     V      &abs-weak&   0    \\
 00032273-2953505 & 12.976 & 12.148  &   -   & 16.62  &  -     & 28.40                &    26    & M3-4  &     V      &  em        &  1 (FS4)  \\ 
 00032487-2953151 & 12.766 & 11.932  &   -   & 15.34  &  -     &   -                  &    0     & K7-M0 &     V      &  abs       &   0    \\
 00025965-2952522 & 12.363 & 11.534  &   -   & 15.85  &  -     &   -                &    0     & M1-3  &     V      &  abs       &   0    \\
  \hline
   quadrant 3\\
  \hline
00024178-2958531  & 12.441 & 11.598  &   -   & 15.60  & 1.34   & 28.51                &    0     & M0-1 &      V     &no H$\alpha$&   0     \\
00021672-2957256\footnotemark[1]  & 14.712 & 14.279  & 20.08 & 15.77  &  -     &   -                  &    0     & K2-4  &    III/V   &    abs     &   0     \\
00022848-2956218  & 13.632 & 12.832  & 22.42 & 17.28  &  -     &   -                  &    0     & M1-3  &      V     &    abs-weak &   0     \\
00021972-2956074  & 11.852 & 11.047  & 21.41 & 14.26  & 1.17   & 29.46                &    81    & M0-1  &      V     &    em-weak &   0     \\
00022589-2952392  & 13.268 & 12.386  &   -   & 16.44  & 1.45   & 29.08                &    73    & M3-4 &      V     &    em      &   0     \\
   \hline
   quadrant 4\\
   \hline
00023679-3007108  & 12.894 & 12.487  & 19.94 & 14.30  & 0.81   &   -                  &    0     &  K3-5 &    III/V   &     abs    &   0     \\
00023545-3007019  & 11.040 & 10.561  & 18.57 & 12.52  & 0.75   & 29.46                &    89    &  K4-5 &      V     &     abs    &   0     \\
00021455-3006443  & 13.672 & 12.836  &   -   & 17.16  &   -    &   -                  &    0     &  M1-3 &      V     &  abs-weak  &   0     \\
00024292-3006323\footnotemark[1]  & 14.998 & 14.579  & 21.14 & 16.20  &   -    &   -                  &    0     &  K3-5 &    III/V   &     abs    &   0     \\
00022427-3006170\footnotemark[1]  & 15.276 & 14.806  & 21.70 & 16.39  &   -    &   -                  &    0     &  K4-5 &    III/V   &     abs    &   0     \\
00022853-3005457  & 14.164 & 13.430  &   -   & 17.52  &   -    &   -                  &    1     &  M1-3 &      V     &  abs-weak  &   0     \\
00023482-3005255  & 11.800 & 11.155  & 20.73 & 13.80  & 0.9    & 30.00                &    85    &  K5-7 &      V     &  em-weak   &   0     \\
00022819-3004435  & 11.140 & 10.510  & 20.20 & 13.04  & 0.96   & 29.76                &    76    &  K4-5 &      V     &   abs-weak &   0     \\
00022512-3004235  & 14.431 & 13.473  &   -   & 18.86  &  -     & 28.63                &    32    &  M3-5 &      V     &     em     &   0     \\
00022289-3002532  & 12.989 & 12.169  & 22.07 & 16.32  & -      & 29.11                &    68    &  M3-4 &      V     &     em     &   0     \\
00024051-3001598\footnotemark[1]  & 12.925 & 12.587  & 19.50 & 14.15  &  -     &   -                  &    0     &  K0-3 &    III/V   &     abs    &  1 (FS3)   \\
00021456-3001115  & 12.233 & 11.467  & 22.04 & 14.60  & 1.24   &   -                  &    0     &  K4-7 &      V     &     abs    &   0     \\

\hline
\end{tabular}
\end{minipage}
\end{table*}
\subsection{Data preparation}
\subsubsection{Target star selection}
The target field in Blanco-1 was selected to include the highest number of stars with a high membership probability ($>$66\%), an apparent magnitude of V$\le$18 and a high X-ray luminosity (log$L_{X}$ $>$ 28~erg~s$^{-1}$). 
We adopted the X-ray luminosities from \citet{Pillitteri2003} and the membership probabilities from \citet{Mermilliod2008} and \citet{Platais2011}. Furthermore, we used results from \citet{Moraux2007}, who investigated the low-mass population of Blanco-1, and \citet{PanagiOdell1997} who performed a photometric and spectroscopic study of stars in the Blanco-1 area. The VIMOS FoV of the cluster area containing the highest number of stars satisfying our selection criteria is centered at RA=0.708292$^{\circ}$ and Dec=-30.0015$^{\circ}$ (see Fig.~\ref{FoV}). The exposure time was selected according to our S/N limit of 20. This S/N limit enabled a rather high temporal resolution of about four minutes including read-out time and still is sensitive enough to resolve the flux of massive CMEs (see section \ref{sectionCMErate}). \\
\subsubsection{Pre-imaging and mask preparation}
For pre-imaging we used the R-filter and three different exposure times (1, 2, and 5 seconds). For the slit assignment we used the VIMOS Mask Preparation Software (VMMPS) as recommended by ESO. Because we selected the HR Orange grism for the science observations it was allowed to place only one slit in dispersion direction. Therefore we lost several of the pre-selected target stars due to slit overlap. Thereby generated empty regions were filled with randomly chosen objects.
\subsubsection{Data reduction}
The data reduction was accomplished with ESOREX. We used the standard scripts \textit{vmmoscalib} and \textit{vmmosscience} which perform step by step calibration of the raw data frames. We performed quality checks on the calibrated science products produced by the pipeline. Checks on the products generated by \textit{vmmoscalib} include detection and tracing of spectra as well as wavelength calibration. The checks on the products generated by \textit{vmmosscience} concern wavelength calibration, sky subtraction, and object spectra extraction. In Table~\ref{bigtable} we list all stars with a sufficient signal to noise ratio (S/N), to perform an investigation of spectral variability. 

\section{Results}
\subsection{Spectral classification}
\label{spectral_classification}
To estimate spectral types for all of our target stars we performed our own spectral classification. We do this in two steps: at first we estimate the spectral types by visually comparing the normalized target spectra to standard spectra \citep[Miles stellar library, ][]{Sanchez-Blazquez2006}. 
As a second step which allows us to distinguish giants from dwarfs we adopt the method of \citet{Prisinzano2012} who used the spectral line ratios of the Ca\,{\sc i} triplet components (6102, 6122, 6162\AA) to the neighbouring Fe\,{\sc i} spectral line (6137\AA). The Ca\,{\sc i} triplet is strong in giant stars and the spectral line wings are sensitive to surface gravity. We used all dwarf and giant star spectra of the Miles library to compute the following line ratios: Ca\,{\sc i}(6102\AA)/Fe\,{\sc i}(6137\AA), Ca\,{\sc i}(6122\AA)/Fe\,{\sc i}(6137\AA), Ca\,{\sc i}(6162\AA)/Fe\,{\sc i}(6137\AA). 
Late-type main-sequence stars show a typical trend when plotting the line ratios over spectral type \citep[cf. Fig.~2 in][]{Prisinzano2012}. From late-G to M stars the line ratios for main-sequence stars are lower than for giant stars.
Giant stars show an upward trend which flattens and then slowly decreases, whereas late-type main-sequence stars show a general downward trend. This behaviour can be seen in all three ratios. 
According to these trends we separate giant and main-sequence stars. We find that the behaviour of the Ca\,{\sc i}(6102\AA)/Fe\,{\sc i}(6137\AA) and the Ca\,{\sc i}(6122\AA)/Fe\,{\sc i}(6137\AA) line ratios of the Blanco-1 stars follow the branch of main-sequence stars.\\
The behaviour of the third line ratio (Ca\,{\sc i}(6162\AA)/Fe\,{\sc i}(6137\AA)) of the target stars follows only partly the behaviour of main-sequence stars. Eight target stars do not correspond to main-sequence stars with respect to their values of the Ca\,{\sc i}(6162\AA)/Fe\,{\sc i}(6137\AA) line ratio. All of them are of spectral type K. All stars of later spectral type show a main-sequence trend. If we take a closer look at the spectra of those eight stars we see that the Ca\,{\sc i} components as well as the Fe\,{\sc i} spectral line are weaker absorption lines as compared to remaining stars. Typically giant stars exhibit strong Ca\,{\sc i} lines, therefore we suggest that these eight stars are more likely main-sequence stars, in agreement with the first two line ratios.\\
From the spectral classification we find 14 K, 12 M stars, and 2 intermediate stars (K7-M0). Eight stars are probably but not clearly confirmed main-sequence stars (see Table~\ref{bigtable}).
\subsection{Membership probabilities}
To select members of Blanco-1 we used kinematic membership probabilities from \citet{Platais2011}. 
\begin{figure}
\begin{center}
\includegraphics[width=7cm]{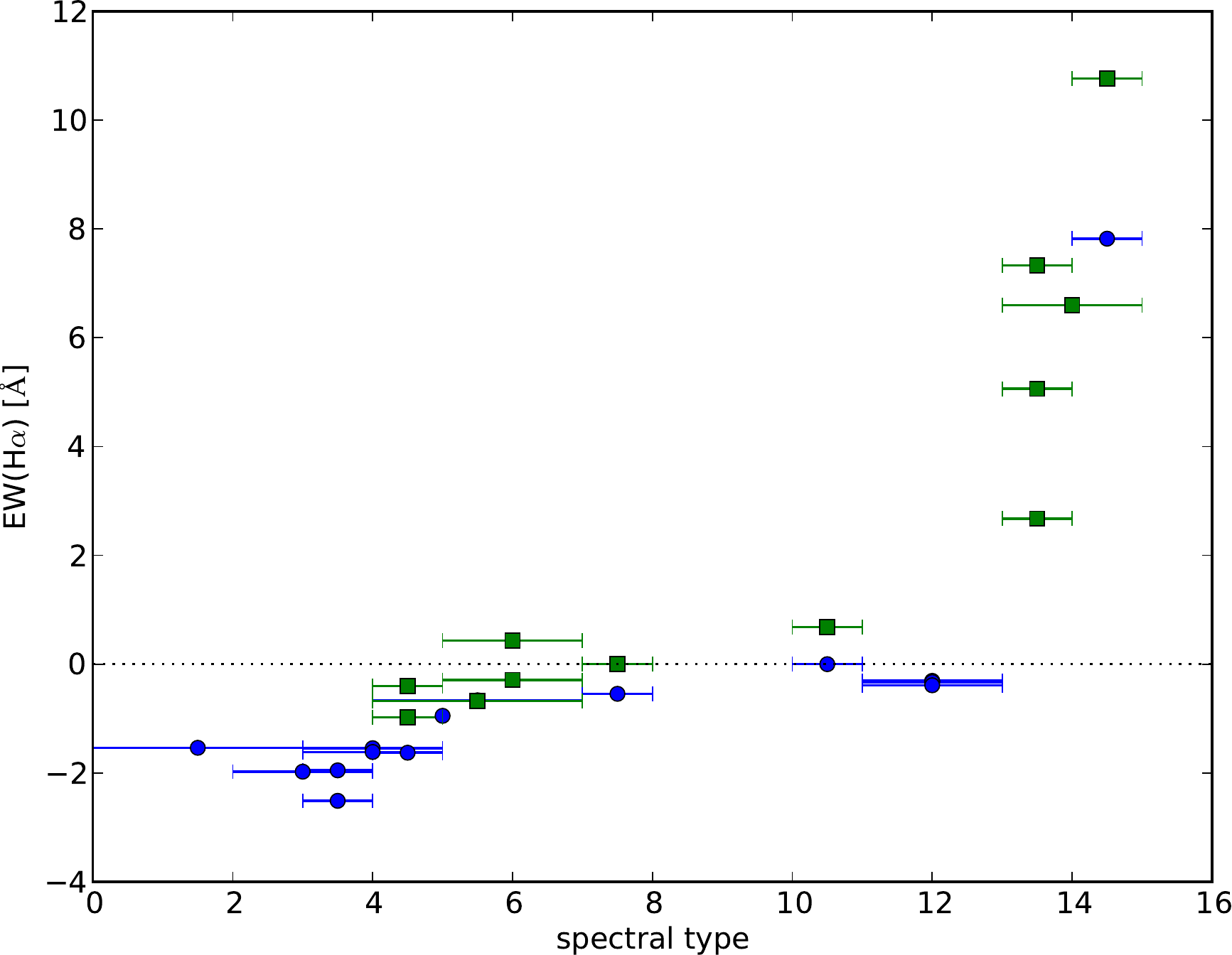}
\end{center}
 \caption{
 H$\alpha$ equivalent width (EW) as a function of spectral type. X-axis labelling corresponds to spectral sub-types K0-M6 (K0=0, M0=10). Negative values of EW denote absorption and positive values emission. The blue filled circles correspond to unlikely member stars (membership probability $<$5\%) whereas the green filled squares correspond to member stars. 
The error bars correspond to the uncertainty in spectral type (see Table~\ref{bigtable}). The statistical errors for the EW(H$\alpha$) are below 1\% for each star, because the EWs were computed from the averaged H$\alpha$ profiles (averaged over the complete time series).}
\label{nuvplot}
\end{figure} 
According to these probabilities 12 stars of the sample are likely cluster members, two stars are possible cluster members, and 14 stars are not cluster members (see Table~\ref{bigtable}). To gain additional information on cluster membership we looked for Ultraviolet (UV) measurements of our target stars. For 16 of the stars we found near-UV (NUV) sources in the database (data release GR7) of the Galaxy Evolution Explorer (GALEX) within a search radius of 10 arcsec around the 2MASS positions. None of the stars were detected in the FUV band.\\
Based on Fig.~7 and Eq.~9 in \citet{Findeisen2011} describing distinct sequences for stars of different ages, we were able to identify inactive stars to support the kinematic non-membership of our sample. All NUV detected sample members are consistent with stars of the age of Blanco-1 or younger.\\ 
Another option to gain information on the activity of the sample stars is the typical activity lifetime of stars as a function of age and spectral type. This was done by plotting the spectral type versus the equivalent width of H$\alpha$ of all sample stars. As one can see from Fig.~\ref{nuvplot} a trend can be found for the equivalent width of H$\alpha$ with respect to spectral type. The later the spectral type the higher the fraction of H$\alpha$ emitting stars \citep[cf.][]{Gizis2002, West2008}. Furthermore, this diagram shows that all kinematically confirmed cluster members are on average more active than the unlikely members of similar spectral type. FS1 is flagged as possible member according to its low kinematic membership of 3\% but from Fig~\ref{nuvplot} we can see that this star (the only blue filled circle symbol at high EWs) fits very well to the other stars of similar spectral type which are kinematically confirmed members. Therefore it is highly likely that FS1 is a member of Blanco-1. From the relation given in \citet{Gizis2002} which describes at which colour (spectral type) H$\alpha$ switches from emission to absorption as a function of stellar age we estimate that for an age of $\sim$ 100~Myr the limiting spectral type lies between K7 and M0. This trend can be also seen in our data.
\subsection{H$\alpha$ line variability}
\subsubsection{H$\alpha$ line profiles}
\label{Halphaprofiles}
Seven stars from the M-star sample (12 stars) show H$\alpha$ in emission (one weak emission line profile), four in absorption (three weak absorption line profiles), and in one star we could not detect H$\alpha$. The K stars of the sample (14 stars) show H$\alpha$ in absorption (two weak absorption line profiles) except for one star (2MASS J00023482-3005255) which exhibits a weak H$\alpha$ emission profile. Two stars have spectral types K7-M0 from which one shows H$\alpha$ in absorption and one shows no H$\alpha$ line (see Table~\ref{bigtable}). 
One of the stars which show no H$\alpha$ line has a high membership probability indicating that this star must have an age similar to the cluster age. \citet{Westerlund1988} and \citet{Epstein1985} estimated ages of 30 and 50~Myr, respectively,  for Blanco-1. \citet{PanagiOdell1997} estimated a higher age of 90$\pm$25~Myr using H$\alpha$ measurements as age indicator. Even recently, \citet{Cargile2014} find an age of $\sim$146~Myr from two different gyrochronology methods.\\
Late-type stars, especially dM stars without an H$\alpha$ line are named dM(e) or zero H$\alpha$ stars \citep{Byrne1990}. According to \citet{Fosbury1974} and \citet{Mullan1975} the presence of an H$\alpha$ absorption line in dM stars is a signature of a non-radiatively heated chromosphere. \citet{Byrne1990} point out that the absence or weak absorption or emission profile of H$\alpha$ detected in dM stars is a matter of stellar evolution or evolution of chromospheric activity. A missing distinct H$\alpha$ profile represents a transition from an active to inactive dM star thereby filling the intrinsic absorption profile due to an increased number of H$\alpha$ active regions. Another possibility is that these dM stars have no chromospheres.
\subsubsection{Flares}
\label{subflare}
\begin{figure}
\begin{center}
$\begin{array}{cc}
\includegraphics[width=4cm]{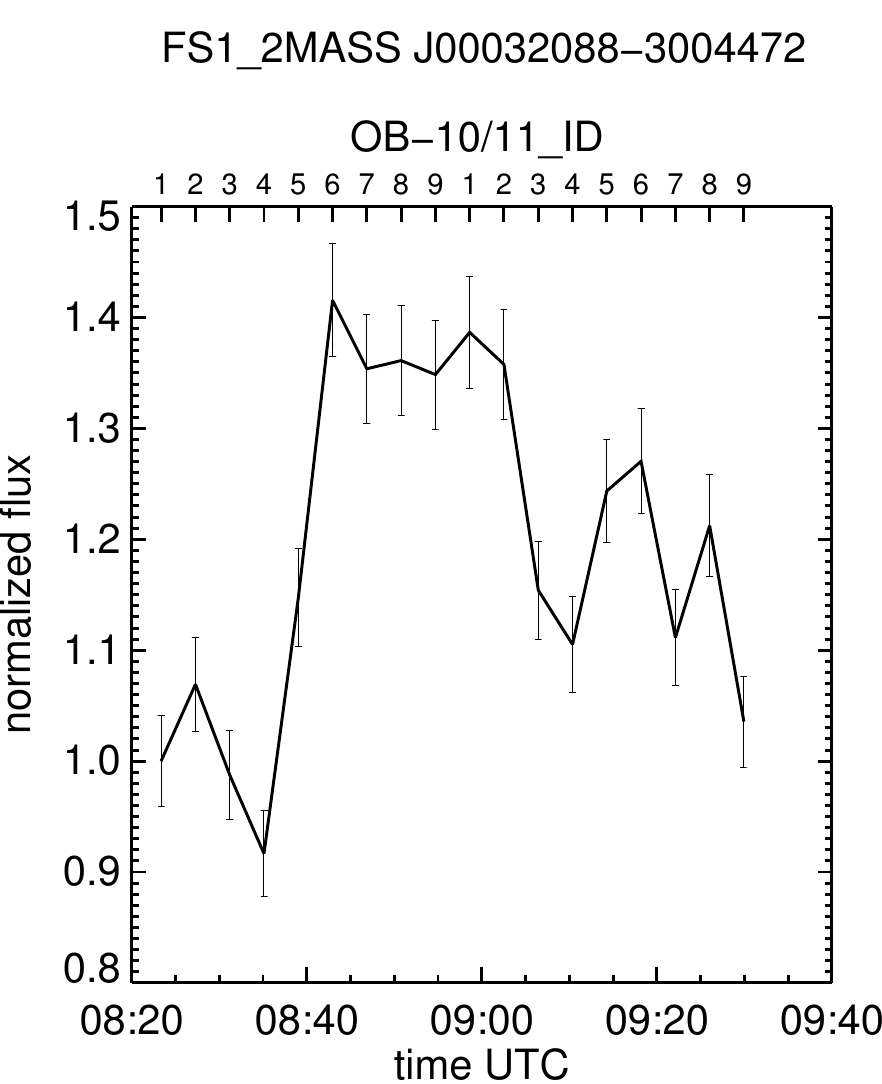} &
\includegraphics[width=4cm]{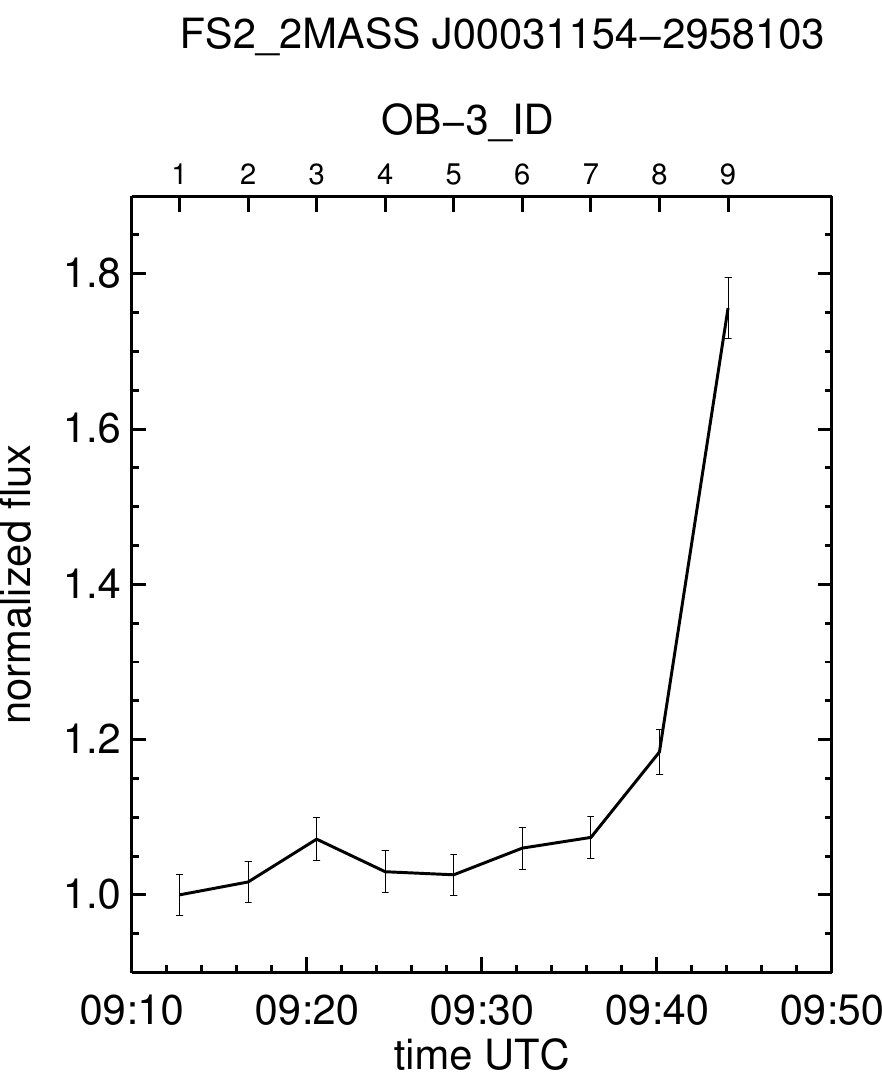}\\
\includegraphics[width=4cm]{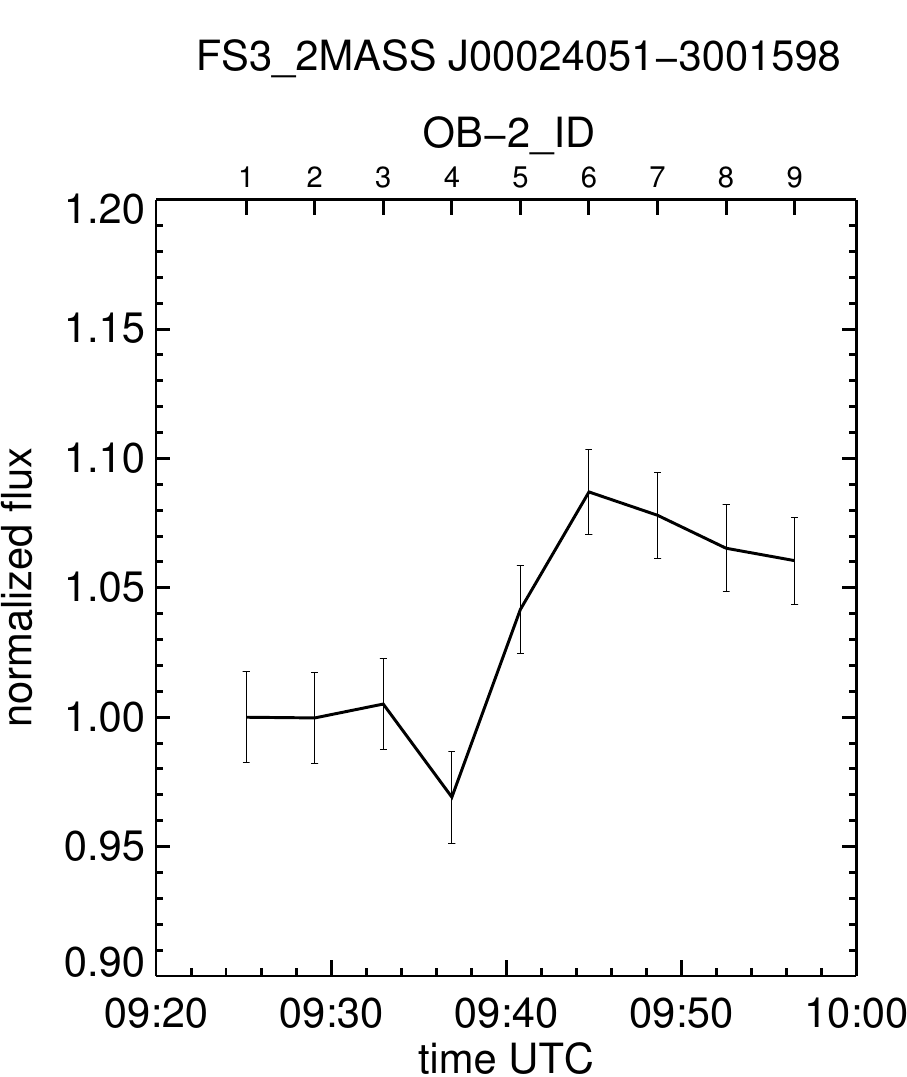} &
\includegraphics[width=4cm]{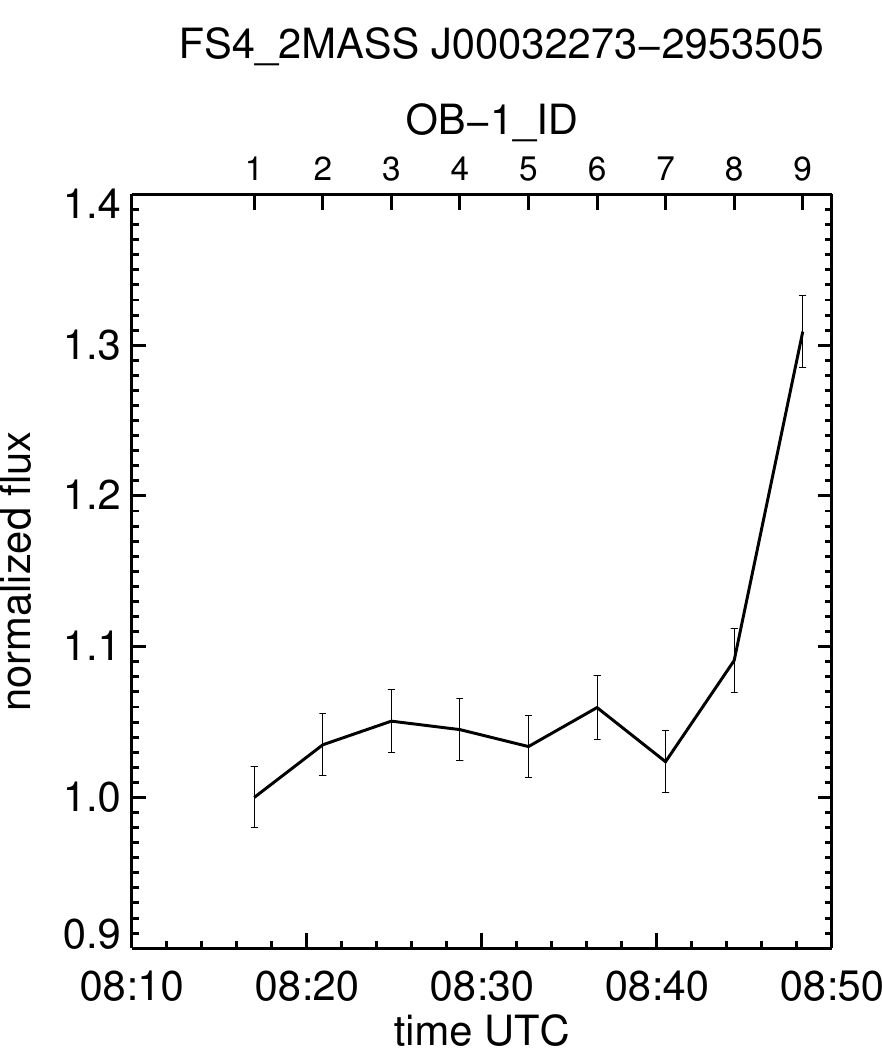}
\end{array}$
\end{center}
\caption{Light-curves of the flares detected on the stars 2MASS J00032088-3004472, 2MASS J00031154-2958103, 2MASS J00024051−3001598, and 2MASS J00032273−2953505. The labelling of the upper x-axis of each plot corresponds to the spectrum numbers of Fig.~\ref{diffplot}. The flux is normalized to the first spectrum of the time series.
}
\label{flarelightcurves}
\end{figure}

To identify flares we calculated the H$\alpha$ line fluxes with a fixed 20\AA{} window centered on H$\alpha$. We found four stars showing distinct flaring activity (see Table~\ref{bigtable}). In Fig.~\ref{flarelightcurves} we show the H$\alpha$ flare light-curves constructed from the spectra. 
\noindent Three additional stars show possible ``fragments'' of flares, either increasing line fluxes at the end of an OB, but less steep than the flares presented in Fig.~\ref{flarelightcurves}, or decreasing line fluxes at the beginning of an OB. The ``flaring stars'' are 2MASS J00032088-3004472 (henceforth termed FS1), 2MASS J00031154-2958103 (FS2), 2MASS J000214051-3001598 (FS3), and 2MASS J00032273-2953505 (FS4). All of the these stars show one flare each. 
To exclude the possibility that the flares are influenced by background emission, we checked the sky spectra for H$\alpha$ emission. In the sky spectra of two of the flaring stars (FS1, FS2) we found weak H$\alpha$ emission. To be sure that the detected sky H$\alpha$ emission does not influence the flares we extracted the H$\alpha$ flux of FS1 and FS2 from the spectra without sky subtraction. This showed that even without sky subtraction the light-curves still have the same shape as with sky subtraction, and therefore the background emission has no influence on the detected flares.
\subsubsection{Flare rates}
To give a parameter on the flare frequency of the stars we computed flare rates and flare duty cycles. For our multi-object observations the flare rate is defined as  
$N_{\mathrm{flares}}/(N_{\mathrm{objects}}\times \Delta t)$.  The flare rate for every flaring star of the study is 0.2 flares per hour (fph). If we do not distinguish between likely (membership probability$>$5\%), possible (0\%$<$membership probability$<$5\%), and non-members (membership probability=0\%) the flare rate for all stars is 0.03 fph. If we take a look at the flare rates for the different member states of the stars then we get 0.03~fph for the likely members (12 stars), 0.1~fph for the possible members (2 stars) and 0.01~fph for the non-members (14 stars). If we consider mid-M type (dM3-5) dwarf stars only (6 stars), we get a flare rate of 0.1~fph.\\
An alternative parameter for the flare frequency is the flare duty cycle which is the ratio of flaring time to total observing time (given in \%) and is defined as $\Delta_ {\mathrm{flare}}/\Delta t$. $\Delta t_{\mathrm{flare}}$ is the number of flare spectra times the duration of one spectrum and $\Delta t$ is the total observing time times the number of stars. For FS1 we get 13\%, for FS2 2\%, for FS3 4\%, and for FS4 we get 2\%. This gives for all stars 0.7\%. For the likely members we get a duty cycle of 0.2\%, for the possible members 3.9\%, and for the non-members we get 0.17\%. For the 6 mid-M-type stars of the sample we calculate 3\%.\\
H$\alpha$ flare rates are rarely reported in the literature. \citet{Lee2010} studied H$\alpha$ variability of M dwarfs and found of 0.1 events per hour with a factor of 2 increase in EW and $<0.05$ per h with a factor 10 increase. \citet{Hilton2010} found from time-resolved Sloan Digital Sky Survey (SDSS) spectra (covering H$\alpha$, H$\beta$) that the flare duty cycle (percentage of time during which flares occur) increases from 0.02\% for early M dwarfs to 3\% for late M dwarfs. If we compare our results to the findings of \citet{Hilton2010} we see that the flare duty cycles of our mid-M-type stars fit well to the 3\% for late-type M dwarfs.
\subsubsection{Flare energies}
To maximize the net observing time per OB we did not request standard star observations for flux calibration. Therefore we need a different method to obtain flare energies. To estimate the total energy output of the flares in H$\alpha$ we calculate the continuum flux using the relations (Eqs. 2 and 3) from \citet{Rutten1989}, using $(R-I)$ colors. As we have no $R$ magnitudes for our target stars we used $(V-I)$ colors and converted to $(R-I)$ colors using the relations given in \citet{Caldwell1993}. In this way we are able to estimate continuum fluxes. The continuum fluxes determined this way were then multiplied with the normalized spectra, and then subtracted from the H$\alpha$ line fluxes. The resulting H$\alpha$ flare energies are 8.95$\times$10$^{28}$ erg for FS1,  $>$2.09$\times$10$^{29}$ erg for FS2, and $>$8.99$\times$10$^{29}$ erg for FS4. The flare energy for FS3 could not be computed due to missing $V$ and $I$ magnitudes. For FS1 the flare energy was computed for the total flaring time. For FS2 and FS4 the flare energies represent lower limits, as only the impulsive flare phases were covered.
\onecolumn
\begin{figure}
\begin{center}
\vspace*{2cm}
$\begin{array}{cc}
\includegraphics[width=8.7cm]{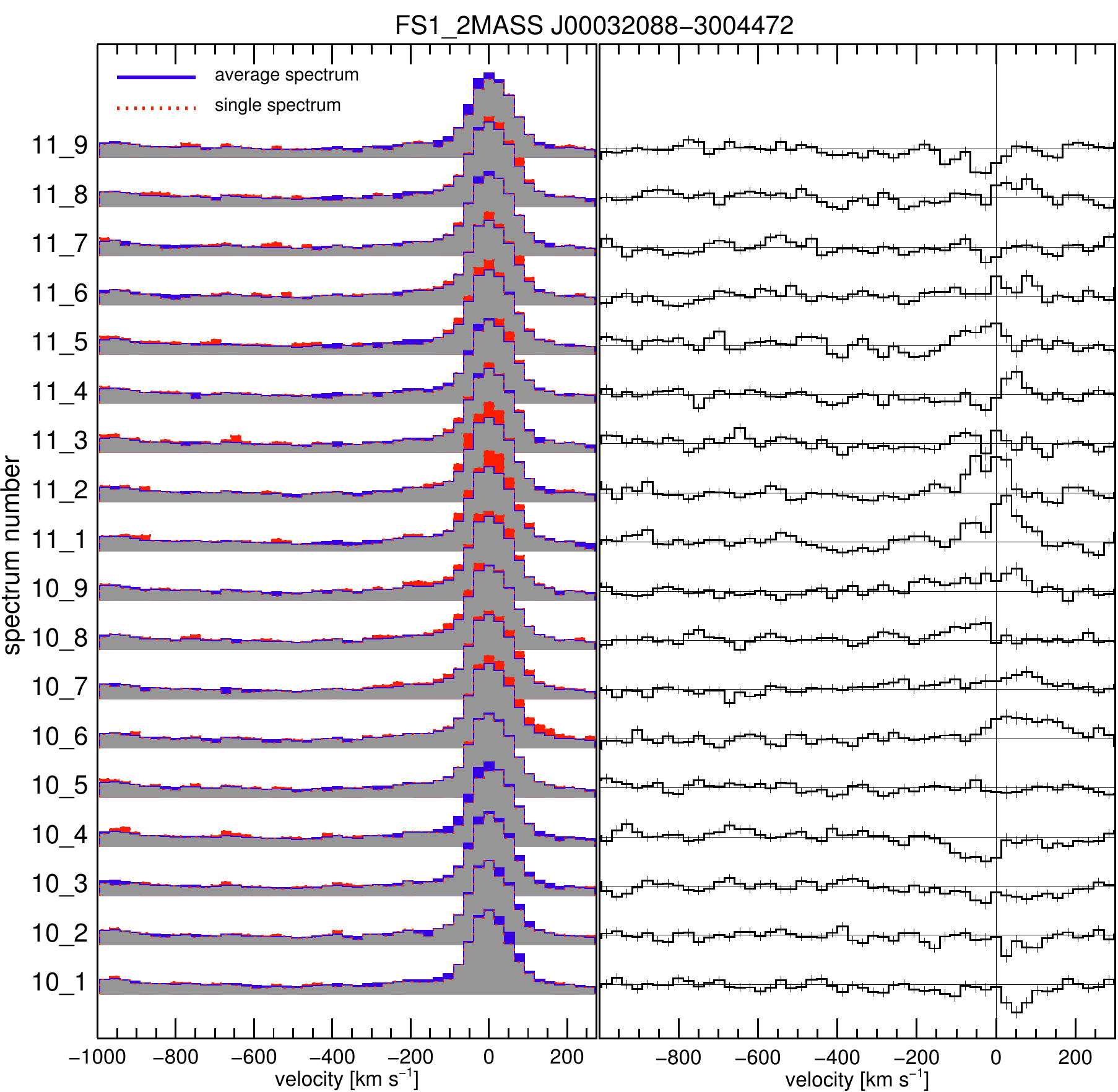}&
\includegraphics[width=8.7cm]{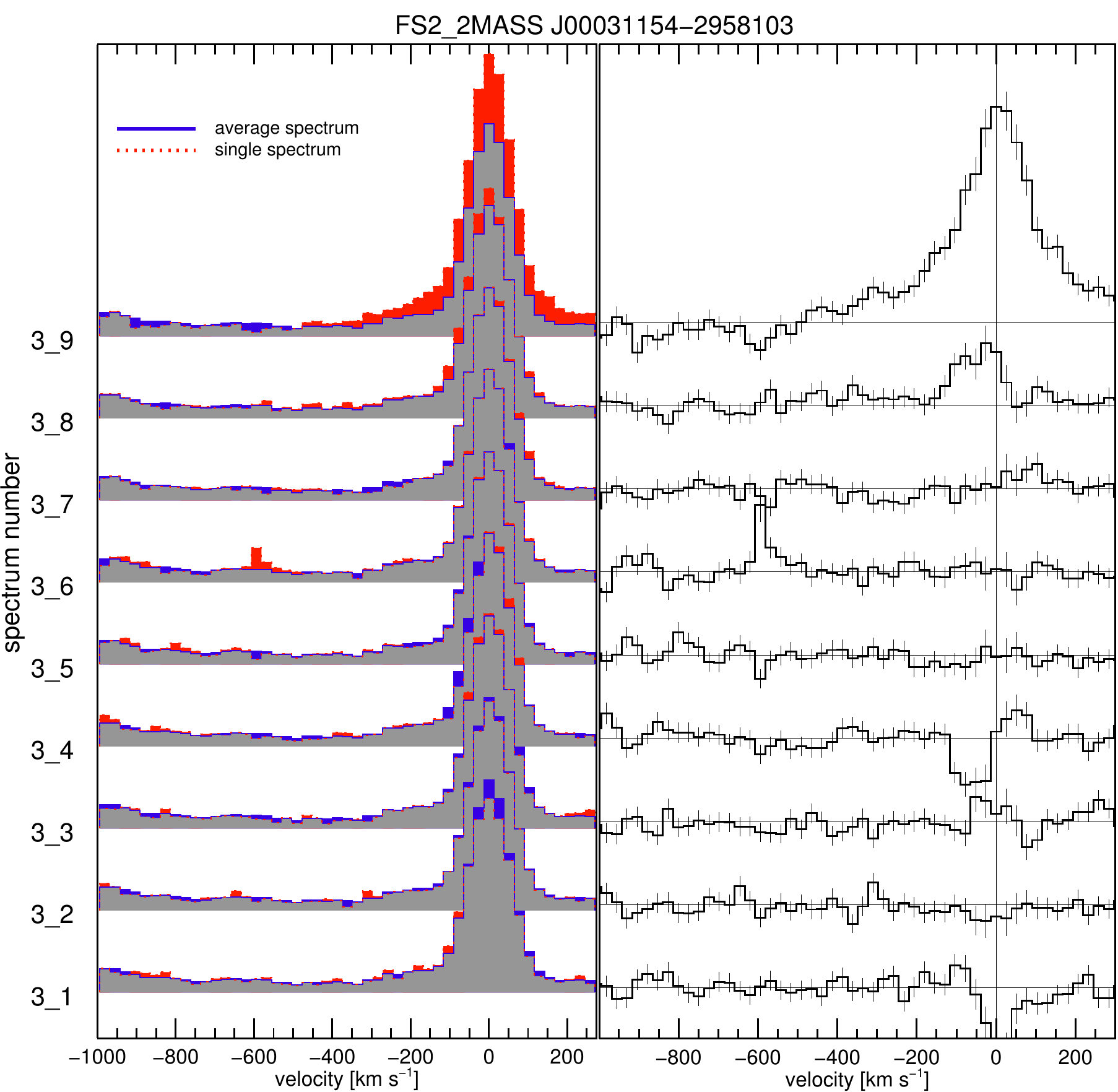}\\
\includegraphics[width=8.7cm]{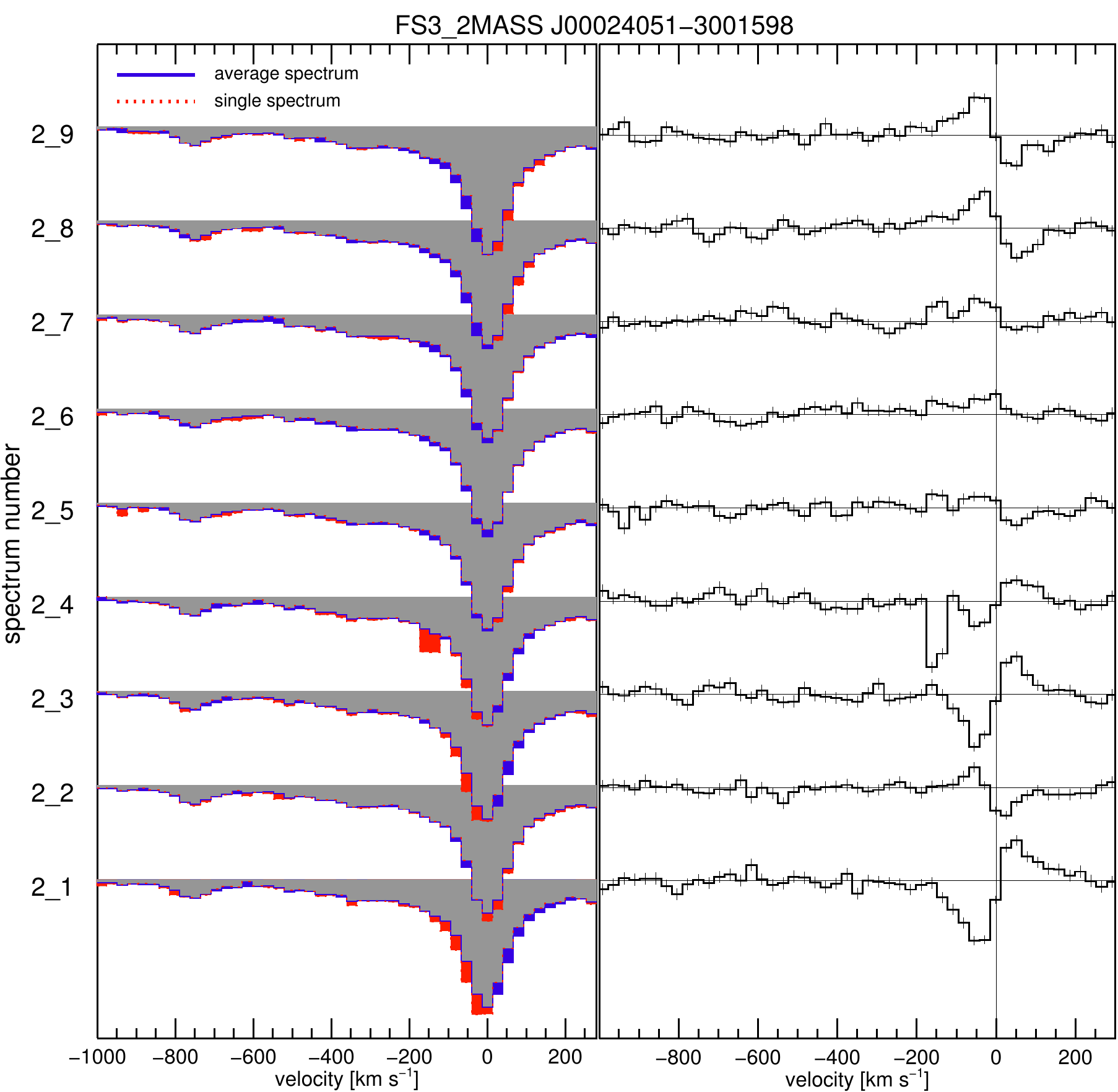}&
\includegraphics[width=8.7cm]{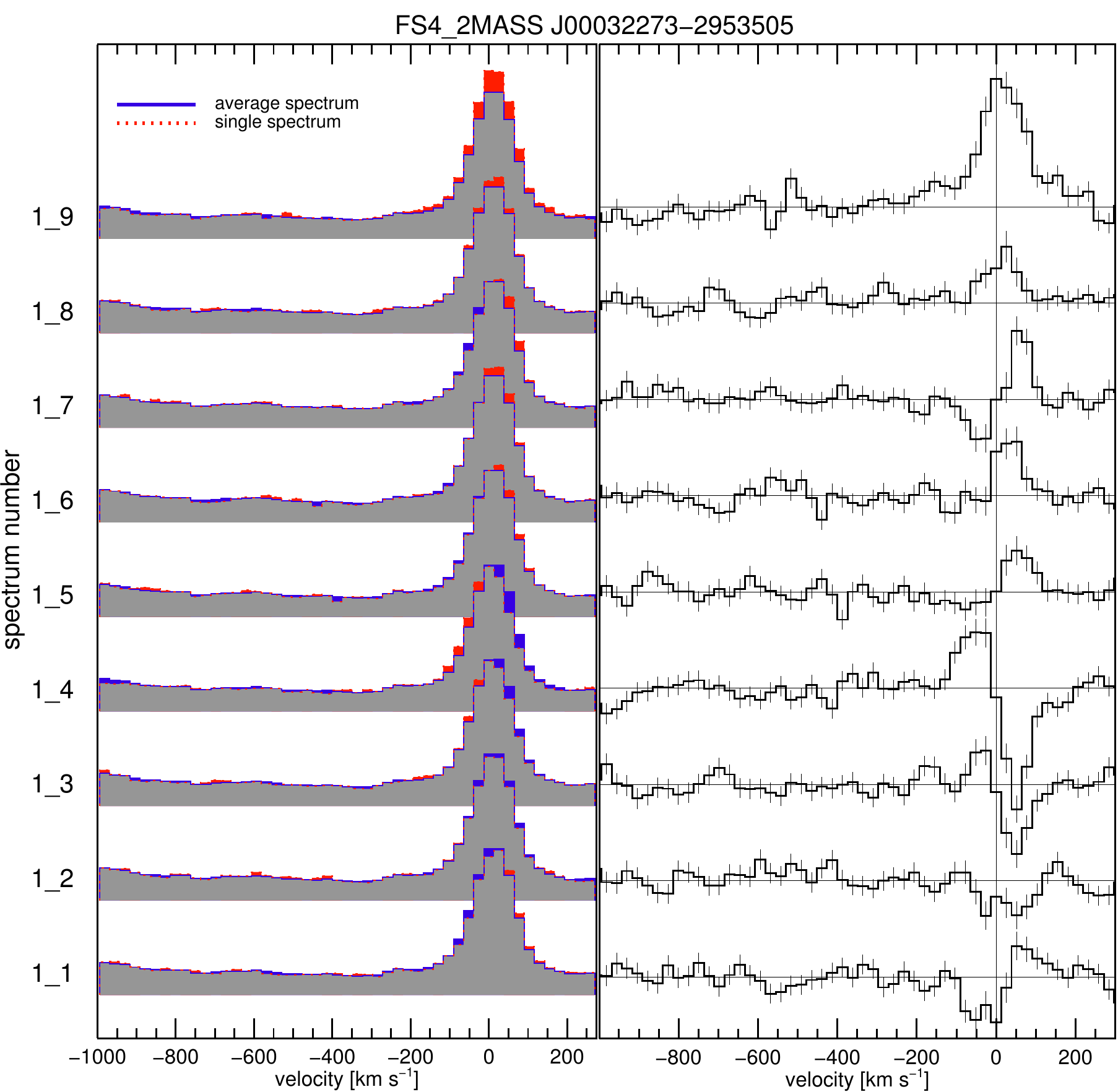}
\end{array}$
\end{center}
\caption{Spectral sequence of H$\alpha$ of FS1 (upper left panel), FS2 (upper right panel), FS3 (lower left panel), and FS4 (lower right panel). In each panel we plot the spectral time series of the flare events from Fig.~\ref{flarelightcurves}. The blue solid line corresponds to the average spectrum of the whole OB, whereas the red dotted line corresponds to the respective individual spectra. Grey areas denote parts of the spectrum which are covered by both average and single spectrum, blue areas correspond to the average spectrum only and red areas to the respective individual spectra. This makes it easily visible where the respective spectrum is lower than the average spectrum (blue) and where the respective spectrum is enhanced (red) with respect to the average spectrum. In the right part of each panel we show difference plots of the individual spectra minus the average one.}
\label{diffplot}
\end{figure}
\twocolumn
    


\subsubsection{Line asymmetries}
The search for line asymmetries was performed by eye. Each OB was analysed separately, because the OBs were not executed in a row. Mainly the M stars in the sample with H$\alpha$ in emission show H$\alpha$ variability. The findings for the ``flaring stars'' are presented in section~\ref{individual} and shown in Fig.~\ref{diffplot}. We did not find any significant H$\alpha$ profile variability in the non-flaring stars.\\
The detectability of stellar CMEs, using the method of Doppler-shifted flux caused by CMEs, depends strongly on the used observational setup. First, the spectral resolution of the spectrograph limits the detectable projected velocity of stellar CMEs. In case of VLT/VIMOS in MOS mode using the HR Orange grism the resolving power is R=2240 in the vicinity of H$\alpha$ which corresponds to $\sim$135~km~s$^{-1}$. This means that stellar CMEs with projected velocities below this threshold cannot be detected.\\
Second, the S/N of the spectra limits the detection of the amount of flux caused by the CMEs. For instance, for FS1 the mean S/N of spectrum 10\textunderscore 1 is 26.9. We calculate a minimum flux of 1.4$\times$10$^{-16}$ erg~s$^{-1}$~cm$^{-2}$ for a stellar CME to be detected in this spectrum with $>$3$\sigma$. For this calculation we used five wavelength bins (ranging from 6555.2-6557.4\AA). The same bins are used in the calculation of the blue wing flux for the individual stars (see section~\ref{individual}). \\
\subsection{Individual stars}
\label{individual}
%
%
%
%
\textbf{\textit{2MASS J00032088-3004472 (FS1):}} 
The flare of FS1 is well covered because OBs 10 and 11 were executed consecutively. In OB 10 the impulsive phase and in OB 11 the gradual decay flare phase are covered. The enhancement of the flare peak with respect to the pre-flare level is $\sim$ 40\%.  In the first four spectra (10\textunderscore 1 to 10\textunderscore 4) we see H$\alpha$ profiles with decreased flux, starting in the red wing (10\textunderscore 1), proceeding to a decrease of the whole profile (10\textunderscore 3), and changing over to the blue wing of H$\alpha$. In spectrum 10\textunderscore 5 the profile is shaped like the average profile until the flux of the core and the red wing of H$\alpha$ significantly rise (spectrum 10\textunderscore 6) which corresponds to the flare peak in the light curve (see Fig.~\ref{flarelightcurves}). In the following spectra (10\textunderscore 7 to 11\textunderscore 9) we detect further H$\alpha$ variability with line-of-sight velocities  of $\sim$ 300~km~s$^{-1}$ until the flux of the H$\alpha$ line approaches its quiescent level. In the gradual phase the H$\alpha$ flux shows three more maxima. In spectrum 11\textunderscore 1 (first maximum during the gradual decay phase) the H$\alpha$ profile is symmetrically enhanced, in spectra 11\textunderscore 6 and 11\textunderscore 8 the flux of the red wing of the H$\alpha$ line is enhanced.\\
The increase of flux in the blue wing of H$\alpha$ relative to the average spectrum as detected in spectra 10\textunderscore 7, 10\textunderscore 8, 10\textunderscore 9 lies in the range of 1-3$\sigma$. Therefore these flux enhancements are not significant. \\ 
Weaker H$\alpha$ profiles in the beginning of the time series are evident in the light curve, forming a ``dip'' shortly before the main rise-phase of the flare. Similar ``dips'' were already reported in the literature. For further discussion we refer the reader to section~\ref{dips_discussion}.\\
%
%
%
%
\textbf{\textit{2MASS J00031154-2958103 (FS2):}} 
The flare of FS2 is the strongest flare of the study with a peak enhancement relative to the quiescent level of nearly 80\%. The flare could not be completely covered due to the limited duration of the OB. At the end of the light-curve one can clearly see a steep flux increase which comes from a sudden overall increase of the H$\alpha$ profile (cf. Fig.~\ref{diffplot}). An inspection of the continua of the spectra showed an overall increased continuum caused by the flare. 
The OB starts with spectrum 3\textunderscore 1 which shows decreased flux in the line core. The following spectra (3\textunderscore 2-3\textunderscore 7) show a rather constant behaviour, except spectrum 3\textunderscore 4 in which the profile exhibits decreased flux in the blue wing of H$\alpha$. In spectrum 3\textunderscore 8 the impulsive phase starts and we detect a distinct H$\alpha$ blue wing enhancement reaching line-of-sight velocities of $\sim$ 175~km~s$^{-1}$. We see increased flux in the line core and in the blue wing. Finally, in spectrum 3\textunderscore 9 the line core as well as both wings are considerably enhanced (line-of-sight velocities of $\sim$ 450~km~s$^{-1}$).\\
The increased flux relative to the average spectrum of the blue wing in the last two spectra of the OB is below the threshold of 3$\sigma$. 
%
\\
%
%
%
%
\textbf{\textit{2MASS J00024051-3001598 (FS3):}} 
The flare of FS3 was not covered completely. We see a flux enhancement at the flare peak of only $\sim$9\% with respect to the quiescent flux level and the flare light-curve shows a typical flare profile.\\ 
In the first four spectra in Fig.~\ref{diffplot} (2\textunderscore 1-2\textunderscore 4) we see decreased flux in the blue wing of H$\alpha$ and enhanced flux in the red wing, except for spectrum 2\textunderscore 2 where the situation is contrary. In the remaining five spectra (2\textunderscore 5-2\textunderscore 9) we find decreased flux in the red wing and enhanced flux in the blue wing, which increase in amplitude towards the end of the OB. The line-of-sight velocities of the blue wing asymmetries reach values of $\sim$ 200~km~s$^{-1}$.\\
%
%
%
%
\textbf{\textit{2MASS J00032273-2953505 (FS4):}} 
The light-curve of the flare of FS4 is similar in shape to the one of FS2, but shows a ``dip'' prior to the impulsive phase of the flare. The enhancement of the flare peak with respect to the quiescent flux level is $\sim$30\%. The light-curve shows pre-flare activity. The first two line profiles of the time series (1\textunderscore 1 and 1\textunderscore 2) have roughly a symmetrical shape. Spectra 1\textunderscore 3 and 1\textunderscore 4 show distinct asymmetric profiles with enhanced blue wings and decreased red wings. This situation is reversed in the next three spectra (1\textunderscore 5, 1\textunderscore 6, 1\textunderscore 7) in which we find enhanced red wings and decreased blue wings. The last two spectra (1\textunderscore 8, 1\textunderscore 9) show an overall enhanced H$\alpha$ profile. The line-of-sight velocities of the blue wing asymmetries reach values of $\sim$ 200~km~s$^{-1}$.\\
If we calculate again the flux increase of the blue wing relative to the average spectrum, all values are below the threshold of 1$\sigma$.  
This detection is not significant.

\section{Discussion}
\label{Discussion}
In this section we discuss the detected ``dips'' in the flare light-curves and the observed and predicted flare rates of the sample stars. Moreover, we discuss the strength of the H$\alpha$ line with respect to spectral type and age. Finally, we discuss the fact that we detected no significant CME activity in the stars of the young cluster Blanco-1 by introducing a semi-empirical CME rate for the observed cluster stars.

\subsection{Flares}
\label{flares_discussion}
From nearly five hours of monitoring late-type stars in a sub-field of the Blanco-1 cluster we found only four H$\alpha$ flares that have a typical shape of a flare light-curve. \citet{Audard2000} deduced from EUV measurements of dF, dG, dK, and dM stars a power-law (see Eq.~\ref{eq3}) connecting X-ray luminosity ($L_{X}$) to flare occurrence. This is a very useful tool for estimating flare rates of such stars. Only for half of the stars in our sample X-ray luminosities are available, either from literature or calculated from XMM Newton fluxes. According to Table~\ref{bigtable} the logarithmic  X-ray luminosities range from 28.40 to 30.00. Those values correspond to 4-60 flares day$^{-1}$ per star exceeding an energy of 10$^{32}$~erg, which is an energy comparable to the most energetic solar flares. Three of the four flaring stars of the present study were detected in X-rays and have log$L_{X}$ values of 28.40, 28.71 and 29.16~erg~s$^{-1}$. These values correspond to flare rates of $\sim$ 4-10 flares day$^{-1}$, respectively, exceeding an energy of 10$^{32}$~erg. In 4.95 hours of observing we would therefore expect 0.25-2.5 flares, respectively. These values agree well with the observed number of flares in H$\alpha$. For the other stars with known X-ray fluxes we would expect up to 15 flares per star during our observing time. It seems to be surprising that no flares could be detected on these stars. 
 \\
The estimated flare rates deduced from X-ray luminosities correspond to X-ray flares. Those originate from a different mechanism than H$\alpha$ flares. On the Sun, the increase in soft X-ray (SXR) radiation is correlated with flare plasma filling the flare loops, whereas the increase in H$\alpha$ flux is correlated with bright flare foot points, which are heated by electron beams which hit denser chromospheric layers \citep{Wuelser1989}. X-ray flares are linked to processes occurring in the corona whereas H$\alpha$ flares are linked to processes occurring in the chromosphere. Correlations of solar X-ray and H$\alpha$ emission during the impulsive phase of flares were established several decades ago \citep[e.g., ][]{Vorpahl1972, Kurokawa1986}. 
On dMe stars and the Sun a correlation of soft X-ray and H$\gamma$ flux of flares was found \citep{Butler1988, Haisch1989, Butler1993}. Even if a statistical correlation between H$\alpha$ and X-ray emission for stellar flares exists, it remains elusive why we detected flares only on four stars and none in the light-curves of the remaining 24 stars.\\
The dK stars in the sample show H$\alpha$ in absorption (except 2MASS J00023482-3005255) indicating that the chromospheric activity level is already decreased, in contrast to the dMe stars. Therefore, probably fewer energetic H$\alpha$ flares may be detected within the $\sim$5 hours of observing time than for the dMe stars. Because dK stars are brighter than dM stars, flares need to be more luminous on those stars to be detected, as a matter of contrast. This means if a flare is detectable on a dM star it does not necessarily mean that a flare with the same energy is also detectable on a dK star. The reason why we detected flares on only 3 out of 7 dMe stars is probably related to the rather small observing window of 5$\sim$ hours and the splitting into several OBs.
\\
\subsection{Pre-flare dips in lightcurves}
\label{dips_discussion} 
As mentioned in section~\ref{individual} we found ``dips'' in the flare light-curves before the impulsive phases  of FS1 and FS4. ``Negative'' flares have been reported in previous studies. \citet{Rodono1979} report on a decrease of flux in a wavelength band comparable to the Johnson B-band during pre-flare activity on the young and active dMe star YZ CMi. The ``dips'' lifetime was estimated to be 10~s with an amplitude of $\sim$10\%. \citet{Giampapa1982} report on a longer ``dip'' observed in $U$-band on the binary system EQ~Peg consisting of two dMe stars. The duration of the ``dip'' was 2.7~min with an amplitude of 25\%. During this decrease in flux no pre-flare activity was detected. \citet{Doyle1988} report again on a ``dip'' observed on YZ CMi. This time the ``dip'' occurred 20~min before the impulsive phase of the flare, had a duration of $\sim$5~min and was observed in $U$, $B$, $V$, $R$, and $I$. \citet{Peres1993} detected in a wavelength band comparable to the Johnson $B$-band several small ``dips'' in flux with durations of $\sim$ 8~min on the binary star FF~And (consisting of two dMe stars) about 25~min before a large flare. \citet{Ventura1995} observed the binary V1054~Oph in $U$, $B$, and $V$. These authors detected a long-lasting ``dip'', mostly prominent in the $U$-band, with a duration of 36~min and an amplitude of 20\%. These ``dips'' have durations of a few minutes, but differ in time of occurrence with respect to the impulsive phase of the associated flares.\\ 
The flares of FS1 and FS4 of the present study show ``dips'' preceding the flare by one spectrum, i.e. three minutes. The duration of the ``dips'' fits very well to the durations reported in literature. The amplitudes lie between 1 and 2~\% measured with respect to the spectrum prior to the ``dip''. The amplitudes of the ``dips'' are small compared to the amplitudes from literature. However, the ``dips'' from literature were detected in photometric bands meaning that the continuum must have been decreased which is not the case in our study because we measure the flux only in a rather small wavelength window around H$\alpha$.\\
A number of explanations can be found in literature regarding ``dips'' occurring prior to the impulsive phase of flares. \citet{Grinin1976} suggested that an increased H$^{-}$ ion density blocks radiation causing the observed ``dips''. 
The predicted timescales lie in the range of seconds, therefore not suitable to explain ``dips'' of longer durations. Nevertheless, there must be a mechanism which reduces radiation for several minutes prior to a flare. \citet{Giampapa1982} suggest that destabilization of filaments may explain the ``dips''. Solar filaments are pronounced in H$\alpha$. Moreover, stellar prominences, co-rotating with the star, are known to cause absorption features overlayed on the H$\alpha$ profile \citep[cf. ][]{CollierCameron1989a}. In the current study, the H$\alpha$ fluxes are calculated by integrating over the H$\alpha$ profile within a fixed window. If a plasma cloud is ejected from the Sun its H$\alpha$ signature is a spectral absorption feature (when seen against the solar disk) occurring on the blue side of H$\alpha$ shifted by the projected velocity of the ejected cloud. In the calculation of the H$\alpha$ flux we include velocities of $\pm$260~km~s$^{-1}$ within a fixed window. Therefore, ascending plasma clouds from the star are included up to 260~km~s$^{-1}$ in our flux determination. In the majority of the studies reported above, ``dips'' preceding flares have been detected in photometric light-curves with no simultaneous spectral observations. In the study by \citet{Doyle1988} simultaneous spectroscopic observations were available but the authors could not detect the U-band ``dip'' in the light-curve constructed from the spectra. In the present study the light-curves have been extracted from the H$\alpha$ spectra. Therefore we are able to reveal which part of the H$\alpha$ spectral line is responsible for the decreasing flux prior to the flare.\\ 
In the flare of FS1 the ``dip'' in the light-curve detected in spectrum 10\textunderscore 4 originates from an asymmetrically (blue wing) decreased H$\alpha$ profile, as it can be clearly seen from the difference plot (upper left panel in Fig.~\ref{diffplot}). The missing flux in the blue wing of H$\alpha$ (with respect to the average spectrum) might indicate an absorbing cloud moving away from the the star with a maximum line-of-sight velocity of 150~km~s$^{-1}$. However, we do not see any significant  absorption in the following spectrum, neither at higher velocities, what would be expected if the phenomenon causing the asymmetric H$\alpha$ profile in spectrum 10\textunderscore 4 is indeed an erupting filament.\\
The ``dip'' in the light-curve of FS3 is caused by a data artefact. In the lower left panel of Fig.~\ref{diffplot} in spectrum 2\textunderscore 4 a sudden drop in flux, seen in two wavelength bins, in the blue wing of H$\alpha$ is evident. This ``feature'' is not present in the raw data and must be therefore a result of the data reduction process.\\ 
The flare of FS4 also shows a ``dip'' prior to the impulsive flare phase. The ``dip'' is caused by decreased flux in the blue wing of H$\alpha$ whereas the red wing flux is enhanced with respect to the average level as in the previous spectra. There exists a similarity between the spectral signatures of the ``dips'' of FS1 and FS4. In both ``dip'' spectra the flux of the blue wing of H$\alpha$ is significantly decreased, the maximum line-of-sight velocities of the H$\alpha$ blue wing absorptions are 100 and 150~km~s$^{-1}$, respectively, and we do not detect any significant absorption feature in the following spectra.\\
We suggest that the detected ``dips'' are most probably explained in terms of stellar pre-flare activity. As pointed out before, the time scales of the ``dips'' in the present study agree well with the time scales reported in literature obtained from photometry. On the other hand, the ``dips'' of the present study are caused by flux variations in the H$\alpha$ profile in contrast to continuum variations. 
\subsection{Spectral signatures of CMEs}
In the analysed spectral time series of 28 stars in a sub-field in the open cluster Blanco-1 we found no indication of fast CMEs. Balmer signatures of CMEs have been detected on a handful of dMe stars so far \citep[e.g., ][]{Houdebine1990, Guenther1997, FuhrmeisterSchmitt2004}. 
The few examples of stellar mass ejections in literature were detected coinciding with flares. Not every flare is necessarily associated with a CME, but from the Sun we know that nearly every energetic (X-class) flare is correlated with a CME \citep{Yashiro2009}.\\
In the present study we did not detect pronounced H$\alpha$ line asymmetries comparable to the ones presented in \citet{Houdebine1990} or \citet{FuhrmeisterSchmitt2004}. The only flare of the analysis which shows line wing asymmetries is the flare of FS1 (see section~\ref{individual} for a detailed description of the spectra).\\ 
Pronounced H$\alpha$ line asymmetries are likely to be related to CMEs as no other sporadic phenomenon is known to produce velocities of several hundreds of km~s$^{-1}$. One major problem is that CMEs are ejected from the star at arbitrary angles, therefore one observes projected velocities only. This makes the identification and interpretation of CME signatures challenging. For such low-velocity events as shown in Fig.~\ref{diffplot} also other interpretations are possible. From the Sun ``gentle evaporation'' is known to be a low-velocity (10~km~s$^{-1}$ in H$\alpha$) upward motion of chromospheric plasma during flares \citep{Schmieder1987, Berlicki2005}. On stars, chromospheric evaporation is assigned to even higher velocities. \citet{Gunn1994} detected a blue wing enhancement in H$\delta$ during a flare which those authors interpreted as a high-velocity chromospheric evaporation. The measured maximum velocity of the event reached 574~km~s$^{-1}$ with a mean of 270~km~s$^{-1}$. The velocity of the event on FS1 is even lower than for the event reported by \citet{Gunn1994}.\\
 
\subsection{Observed vs. expected CME rate}
\label{sectionCMErate}
Although we did not detect reliable evidence for signatures of stellar CMEs we calculate an upper limit of the CME rate per star. For one star the CME rate is defined as  $N_{\mathrm{CME}}$/$\Delta\tau$, where $N_{\mathrm{CME}}$ is the number of CMEs and $\Delta\tau$ is 4.95~h (0.2~d), which is the length of the observing period. 
We detected no CME, therefore $N_{\mathrm{CME}} < 1$. The upper limit of the CME rate is therefore $<$~4.85 CMEs per day.\\ 
However, if we consider that we did not detect a CME on all simultaneously monitored 28 stars, the total average CME rate is $N_{\mathrm{CME_{tot}}}$/$\Delta\tau$ and therefore must be also $<$~4.85, as for the single star, as explained above. The average upper limit to the CME rate per star for the whole sample is $<$ 0.17 per day. These are limits to the observable CME rate which differs from the intrinsic rate by a detection efficiency factor which depends on instrument parameters and method of detection. \\
The detection of stellar CMEs is limited by two main factors, namely the intrinsic properties of the observed star and observational constraints. The former is the CME occurrence rate, which is related to the activity level/age of the star, and the typical CME parameters (velocity, mass etc.). The latter is related to temporal aspects (time coverage and time resolution), wavelength (i.e. velocity) resolution and S/N of the observations.\\
Young, Sun-like and late-type stars are usually more active and have more frequent and powerful flares. Assuming that the physical processes causing the flares are similar to the Sun, it is likely that young stars should also have a large number of CMEs, because on the Sun these phenomena are related and especially strong flares are almost always accompanied by CMEs. Hence, we estimate the expected CME occurrence rate of our target stars by estimating the frequency of strong flares. On the Sun and stars, the distribution of flares according to their energy follows a power law
\begin{equation}\label{eq1}
\frac{\mathrm{d}N}{\mathrm{d}E}=kE^{-\alpha}
\end{equation}
with index $\alpha$ and normalization constant $k$. Indices $\alpha$ between 1.5 and 2.5 have been found for both solar and stellar flares \citep[e.g.][and references therein]{Audard2000}. The normalization constant $k$ is roughly proportional to the stellar X-ray luminosity $L_{X}$ and is therefore related to the stars activity level. \citet{Audard2000} found that on young stars the daily number of flares with X-ray energies $E>10^{32}$\,erg scales with the stellar X-ray luminosity $L_{X}$ as
\begin{equation}\label{eq3}
\log N(E>10^{32}) = (-26.7\pm2.9)+(0.95\pm0.1)\log L_{X}.
\end{equation}
This energy limit is comparable to the largest solar flares, which are almost always accompanied by CMEs. By comparing Eq.~\ref{eq3} with the cumulative distribution of flares with energies $E>E_C$, where E${_C}$ is a chosen cut-off energy, 
\begin{equation}\label{eq2}
N(E>E_C)=\frac{k}{\alpha-1}E_C^{-\alpha+1} = \tilde{k}E_C^{-\alpha+1},
\end{equation}
one can see that the normalization factor of the cumulative distribution is
\begin{equation}\label{eq4}
\tilde{k}=\frac{k}{\alpha-1}=\frac{N(E>10^{32})}{10^{32(-\alpha+1)}}
\end{equation}
where $N(E>10^{32})$ can be estimated from Eq.~\ref{eq2}.\\
On the Sun CME masses correlate with the X-ray energies of the associated flares as $M_\mathrm{CME}=\mu E_f^\beta$ (E$_{f}$ is the X-ray energy in the 1--8\AA\ GOES channel) with $\mu=10^{-1.5\mp0.5}$ and $\beta=0.59\pm0.02$ \citep{Drake2013} or $\mu=(2.7\pm1.2)\times10^{-3}$ and $\beta=0.63\pm0.04$ \citep{Aarnio2012}. The latter parameters yield up to about a factor of two higher CME masses for a given flare energy. This difference is probably due to the different methods used to calculate the X-ray flare energies. In the calculation of the CME rate we adopt the parameters $\mu$ and $\beta$ from \citet{Drake2013}. Those fits are given for the mean CME mass for a given flare energy, but the intrinsic scatter about this mean value is about an order of magnitude \citep{Aarnio2011}. The X-ray energies used in these solar studies were derived from the 1--8\AA\ GOES channel, which is a narrower energy band than used in the stellar X-ray observations \citep{Audard2000}. We adopt a correction factor $C_{GX}=15.4\pm0.8$ \citep{Emslie2012} to convert between GOES energies and the energy in the total soft X-ray band. Expressing the CME mass--flare energy relation directly as a function of the total soft X-ray energy $E$ thus yields
\begin{equation}\label{eq5}
M = \mu\left(\frac{E}{C_{GX}}\right)^\beta = \tilde{\mu} E^\beta
\end{equation}
with $\tilde{\mu}=\mu/C_{GX}^\beta$. To test if this CME mass--flare energy relation holds for stars, we compare it with the observed flare and CME parameters of the M dwarfs AD~Leo \citep{Rodono1984, Houdebine1990} and AT~Mic \citep{Gunn1994}, as well as the M-type WTTS J1149.8-7850 \citep{Guenther1997}. No X-ray observations are available for these flares, therefore we use the total $U$-band energy as a proxy, which is of similar magnitude as the X-ray energy for M stars \citep{Hawley1991}. The published $U$-band energies of these flares are $\sim2\times10^{32}$\,erg \citep{Rodono1984,Hawley1991}, $\sim3\times10^{31}$\,erg \citep{Gunn1994} and $1.2-2.5\times10^{35}$\,erg \citep{Guenther1997}, and the minimum masses of the associated CMEs were estimated to be $7.7\times10^{17}$\,g, $\sim10^{15}$\,g, and $1.4\times10^{18}-7.8\times10^{19}$\,g. Using Eq.~\ref{eq5}, we calculate CME masses of $7\times10^{16}$\,g, $2\times10^{16}$\,g, and $3-5\times10^{18}$\,g for the given flare energies. These numbers agree to within about an order of magnitude, which is comparable to both the intrinsic spread of solar CME masses for a given flare energy, as well as the quoted uncertainties of the estimates of the minimum stellar CME masses. However, Eq.~\ref{eq5} is expected to be more useful for statistical purposes and might fail for individual flare events.\\
The association between flares and CMEs on the Sun is very high for strong flares, but declines with decreasing flare energy \citep{Yashiro2006, Yashiro2009, Drake2013}. The CME association rate is about one for GOES X-ray flare energies above $3.5\times10^{29}$\,erg \citep{Drake2013}, which corresponds to an average CME mass of about $8.5\times10^{15}$\,g after their relation. For simplicity, we therefore neglect CMEs with lower masses/flares with lower energies in the estimation of stellar CME occurrence rates which is justified by the detection limit of our observations. \\
\begin{figure}
\begin{center}
\vspace*{0cm}
\includegraphics[width=8.5cm]{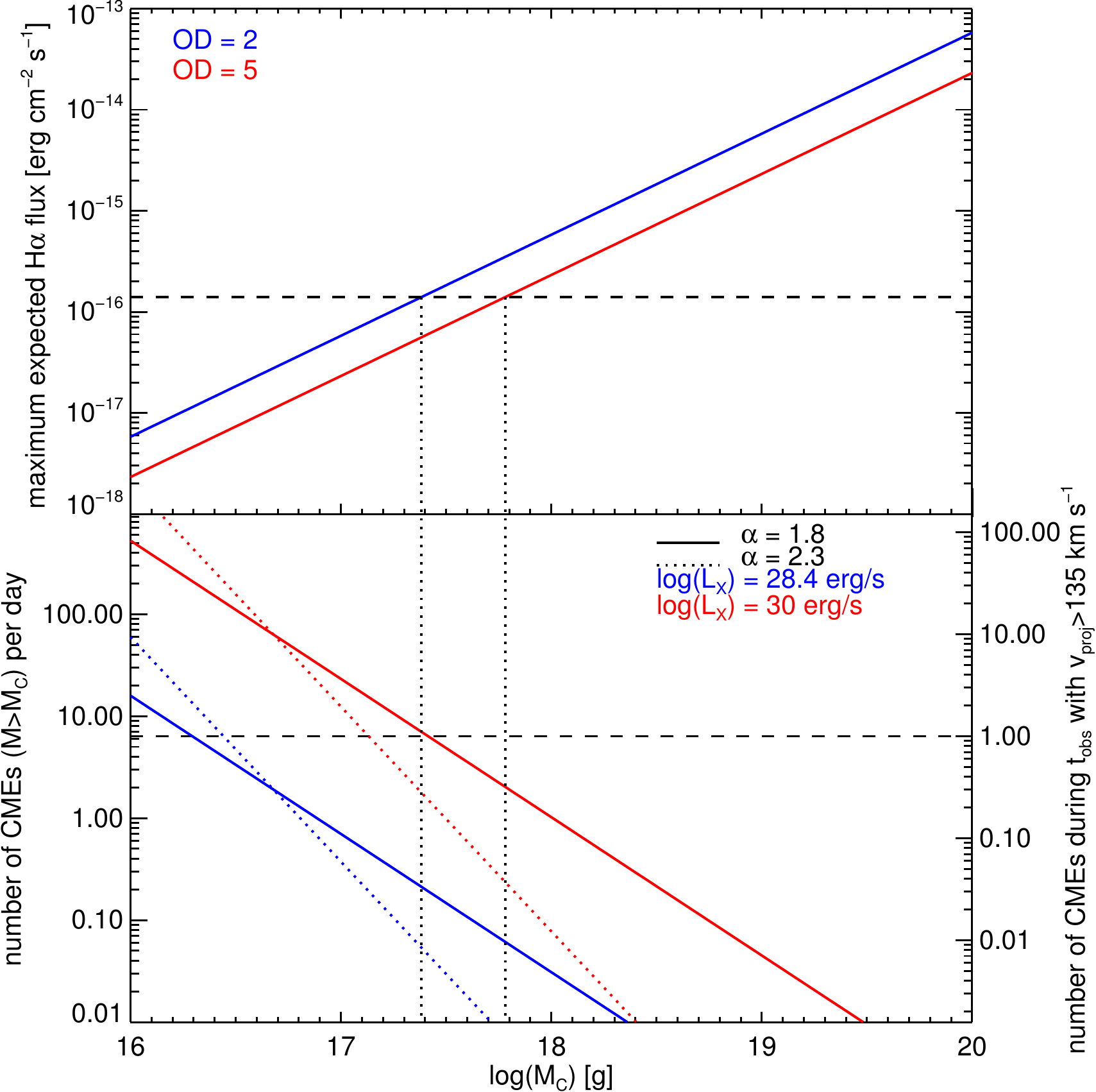}
\end{center}
\caption{Lower panel: Semi-empirical CME rates per day (left ordinate) and per observing time (right ordinate) for two different flare indices and log$L_{X}$ values. Red coloured lines correspond to a log$L_{X}$ value of 30 erg~s$^{-1}$ (highest log$L_{X}$ value of the sample stars) and blue coloured lines correspond to a log$L_{X}$ value of 28.4 erg~s$^{-1}$ (lowest log$L_{X}$ value of the sample stars). Solid lines correspond to a flare index of 1.8 and dotted lines correspond to a flare index of 2.3. Upper panel: Dependence of maximum expected H$\alpha$ flux on CME mass for two different opacity damping factors (red and blue solid lines). The horizontal dashed line marks the detection limit of our observations (a matter of the S/N). The dotted vertical lines indicate which H$\alpha$ flux associated with CME mass we are able to detect with our observations. As one can see we are not able to detect H$\alpha$ flux associated to CMEs with a mass less than 4-8$\times$10$^{17}$g.}
\label{cmerate}
\end{figure}
Using Eqs.~\ref{eq1}--\ref{eq5}, the daily number of CMEs with masses $M>M_C$, where M$_{C}$ is a chosen cut-off CME mass, can then be written as
\begin{align}\label{eq6}
N(M>M_C) &   =  \int_{M_C}^\infty\frac{\mathrm{d}N}{\mathrm{d}M}\mathrm{d}M = \\
	     &   =	\int_{M_C}^\infty\frac{\mathrm{d}N}{\mathrm{d}E}\frac{\mathrm{d}E}{\mathrm{d}M}\mathrm{d}M = \nonumber\\
		 &   =  \tilde{k} \tilde{\mu}^\gamma M_C^{-\gamma} \nonumber 
\end{align}
with $\gamma=(\alpha-1)/\beta$. In Fig.~\ref{cmerate} we show the cumulative number of CMEs per day from Eq.~\ref{eq6} as a function of $M_C$. We show four different cases, two values of the flare power law index $\alpha$ and two values of the stellar X-ray luminosity. The chosen values of $\alpha$ are 1.8 and 2.3, bracketing the average values of F--M dwarfs found by \citet{Audard2000}. The two values of the stellar X-ray luminosity correspond to the minimum and maximum observed values from our target stars (cf. Table~\ref{bigtable}). From Fig.~\ref{cmerate} it may seem that M dwarfs (lower $L_{X}$) may be less suitable targets compared to Sun-like stars of comparable age. However, as most stellar CME detections in literature so far have been reported for young M dwarfs, the detectability may be better for such stars.\\
To estimate the number of CMEs that can be observed we have to apply some corrections. The most obvious correction is that of the net observing time, which is 4.95~h in the present case. Another correction is due to projection effects related to the line-of-sight velocity. If the angle between the direction of ejection and the line-of-sight is large, the CME will be undetected even if its true velocity is high. To estimate the fraction of CMEs lost due to velocity projection effects, we adopt the distribution of CME velocities from the Sun and multiply it with the cosine of a randomly distributed projection angle. The distribution of solar CME velocities follows a lognormal and we adopt a mean value of $\ln v=6.09$ and $\sigma=0.53$, which has been corrected for projection effects of solar observations and thus resembles the distribution of the true solar CME speeds \citep{Wu2011}. From our data we estimate that detectable CMEs should have a minimum projected velocity of $\sim135\,\mathrm{km\,s^{-1}}$. Applying this limit to the distribution of projected velocities, about 20\% of CMEs are lost due to velocity projection. However, if the average CME speed on young stars is larger than on the Sun, this percentage decreases. Note that for velocities of $\sim135\,\mathrm{km\,s^{-1}}$ (spectral resolution of the observational setup of the VLT/VIMOS observations) one cannot unambiguously conclude from the data alone that a CME has occurred, because such shifts can also be due to mass motions with true velocities of this magnitude which would not leave the star because they are lower than the escape speed. On the other hand, if the line-of-sight velocity is already larger than the escape speed of the observed star, then a CME event is likely. The right ordinate in Fig.~\ref{cmerate} gives the number of CMEs expected during the net observing time per star in this study (4.95~h) and is corrected for the velocity projection.\\
Another issue is the time during which a CME feature may remain visible. Here one has to distinguish between emission and absorption features. For absorption features the visibility is limited by the on-disk time, i.e. the ratio of the distance between the location of ejection to the border of the stellar disk in the direction of the ejection to the velocity component in this direction. For a Sun-like star ($R_*=R_{\sun}$, typical CME ejection locations about $\pm30\degr$ around equator, velocity distribution like above) we estimate that about 70\% of all CMEs ejected into random directions have on-disk times longer than three minutes, the exposure time of our spectra. However, emission features can still be seen outside the stellar disk until the CME dissolves. We note that all Balmer signatures related to stellar CMEs in literature, which were predominantly detected on M dwarfs, were emission features, so we conclude that limited on-disk times do not alter our estimated detection rates. With the calculations presented above we refer to CMEs which produce emission-features only.\\
Another limit affecting the detectability of a stellar CME event is the amount of flux modulation caused in the spectra. Although there is no simple relation between the mass of the ejected material and the feature it leaves in the spectrum, because it also depends on plasma parameters which can differ between events, it is nevertheless likely that more massive CMEs should cause stronger flux modulations. This parameter is only roughly constrained by the very few observations of stellar CMEs and their corresponding mass estimates. To get a rough estimate on the expected H$\alpha$ flux for different CME-masses, we follow the method by \citet{Houdebine1990} used to determine the minimum CME-mass from the observed H$\gamma$ flux modulation. Since the observed flux gives a lower limit to the number of atoms in the upper level of H$\gamma$, inversely, an upper limit of the expected observed flux $F_{\mathrm{H}\gamma}$ (erg\,cm\textsuperscript{-2}\,s\textsuperscript{-1}) for a given CME mass can be derived by rearranging their Eq.~3, yielding

\begin{equation}
F_{\mathrm{H}\gamma} \leq \frac{N_5 h \nu_{5-2} A_{5-2}}{4 \pi d^2} = 
    \frac{h \nu_{5-2} A_{5-2}}{4 \pi d^2} \frac{M_C}{(N_\mathrm{tot}/N_5) \bar{m}},
\end{equation}

with the number of atoms in energy level 5 $N_5$, Planck's constant $h$, the H$\gamma$ transition frequency $\nu_{5-2}$, the Einstein coefficient of the transition $A_{5-2}$, the distance to the star $d$, for which we adopt the mean distance to Blanco-1 (240~pc), the true CME-mass $M_C$, the ratio of the total number of hydrogen atoms to atoms in level 5 $N_\mathrm{tot}/N_5 \sim 2\times10^9$ \citep{Houdebine1990}, and the mean particle mass $\bar{m}=1.3m_\mathrm{H}$ for neutral gas with solar abundances. Considering that the H$\gamma$ line is optically thick for typical flare plasma parameters, \citet{Houdebine1990} introduced the opacity damping parameter $OD$, the ratio of the optically thin to the real flux. For instance, $OD=2$ means that only 50\% of the radiation escapes from the plasma, i.e. $F_{\mathrm{H}\gamma,\mathrm{obs}}=F_{\mathrm{H}\gamma}/OD$. We then estimate the expected maximum H$\alpha$ flux from $F_{\mathrm{H}\alpha}=3F_{\mathrm{H}\gamma,\mathrm{obs}}$, the observed Balmer decrement of solar and stellar flares \citep{Butler1988, Guenther1997}. In the upper part of Fig.~\ref{cmerate}, we show the estimated $F_{\mathrm{H}\alpha}$ as a function of CME-mass and compare it to the detection limit flux calculated from our data. It becomes clear that the S/N of our data is likely too low to find clear indications of stellar CMEs during our observing time, because only our most active star could have had one CME-event producing H$\alpha$ emission above our detection limit. Follow-up studies thus require longer observing times and/or higher S/N, the latter being complicated by the fact that relatively short exposure times in the order of a few minutes are necessary to have sufficient temporal resolution to catch these transient events. \\
Moreover, an unknown correction term related to the projected shape of a CME of a given mass would possibly lower the numbers shown in Fig.~\ref{cmerate}. We also ignored possible differences between stellar spectral types in both intrinsic CME occurrence rate and detectability.\\
Taking all aspects discussed above into account and considering the log$L_{X}$ distribution of the sample stars we find that we should have detected at least one CME per star in the 4.95~h of spectral monitoring with masses in the range of $1-15\times$ 10$^{16}$g (cf. Fig.~\ref{cmerate}). However, from our estimated maximum H$\alpha$ fluxes one can see that this CME mass range is below the observational detection limit of the current study.\\
A factor which remains unknown are the activity cycles of the stars, although at the age of Blanco-1 the average activity level should still be high. Moreover, no dependence on spectral type is considered in the empirical estimates. The fact that the total observing time was split into 11 OBs might also influence the result because with this observing strategy one gets 11 snapshots of 27~min each, thereby ignoring stellar activity occurring before and/or after the snapshot. In general, it seems that an observing time of a few hours in connection with a moderate S/N limit is not sufficient to detect signatures of CMEs in H$\alpha$ even for active stars.\\
The CMEs detected in \citet{Houdebine1990} and \citet{Guenther1997} occurred both during impulsive phase of flares. The same is valid for the blue wing enhancement detected in a flare on AT Mic \citep{Gunn1994}, termed by the authors as ``evaporation''. All of these flares showed high energies as described above (10$^{31}$-10$^{32}$ erg for the dMe stars and 10$^{35}$ erg for the WTTS). The H$\alpha$ energies of the flares of the present study lie in a range of $\sim$10$^{29}$-10$^{30}$ erg which is a factor of 10-100 lower than the flare energies of dMe stars from literature. This finding may explain why we have not detected signatures of CMEs. If we consider the correlation of strong flares and CMEs from the Sun to be valid for young dMe stars, then the flares presented in this study were likely too weak to be accompanied by CMEs.\\ 




\section{Conclusion and outlook}
We have undertaken a search for stellar activity in 28 stars of the open cluster Blanco-1. The spectral classification revealed that the majority of the stars are main-sequence stars except eight stars which could not be clearly classified as main-sequence stars. The analysis of the spectral time series showed four flares. In two of the four flares we found a decrease in H$\alpha$ flux prior to the impulsive flare phase, reminiscent of ``dips'' in previous stellar photometric analyses. We did not detect any clear signs of stellar CMEs, but deduce an upper limit of the average CME rate of $<$~0.17 CMEs day$^{-1}$ per star for the sample. We estimate a semi-empirical CME rate, based on distributions of solar CME parameters, for stars with a known X-ray luminosity considering projection effects.
This semi-empirical CME rate predicts at least one CME per star in the observing period within a CME mass range of $1-15\times$ 10$^{16}$g, which is below the detection limit of our observations. We conclude that it needs longer observing times and higher S/N ratios to detect CMEs, even on young and active stars. 

\section{Acknowledgments}
 ML, PO, and AH acknowledge support from the FWF project P22950-N16. HK acknowledges the support from the European Commission under the Marie Curie IEF Programme in FP7. HL and MK acknowledge the support by the FWF NFN project S116 ``Pathways to Habitability: From Disks to Active Stars, Planets and Life'', and the related FWF NFN subprojects, S116 606-N16  and S116607-N16. We thank the anonymous referee for constructive comments.

       \bibliographystyle{mn2e_update}
   \bibliography{Mybibfile}

\end{document}